\documentclass[10pt,nofootinbib,pra,twocolumn]{revtex4-2} 

\setcitestyle{super} 
\usepackage[utf8]{inputenc}
\usepackage{upgreek} 
\usepackage{graphicx} 
\usepackage{hyperref}
 \hypersetup{
     colorlinks=true,
     linkcolor=blue,
     filecolor=blue,
     citecolor = red,      
     urlcolor=cyan,
     }   

\makeatletter
\renewcommand\LARGE{\@setfontsize\LARGE{15pt}{17}}
\renewcommand\Large{\@setfontsize\Large{12pt}{14}}
\renewcommand\large{\@setfontsize\large{10pt}{12}}
\makeatother

\usepackage{fancyhdr}
\begin{document}


\title{\LARGE{Multimodal microscopy for characterization of amyloid-$\upbeta$ plaques biomarkers in animal model of Alzheimer's disease$^{\dag}$}}
\author{Renan Cunha$^{\ddag}$\textit{$^{a}$}}
\author{Lucas Lafeta$^{\ddag}$\textit{$^{a}$}}
\author{Emerson A. Fonseca$^{\ddag}$\textit{$^{a,b}$}}
\author{Alexandre Barbosa\textit{$^{a,c}$} Marco A. Romano-Silva\textit{$^{d}$}}
\author{Rafael Vieira\textit{$^{e}$}}
\author{Ado Jorio\textit{$^{a,b}$}}
\author{Leandro M. Malard\textit{$^{\ast a}$}}


\newcommand\nsfootnote[1]{%
  \begingroup
  \renewcommand\thefootnote{}\footnote{#1}%
  \addtocounter{footnote}{-1}%
  \endgroup
}


\nsfootnote{\textit{$^{a}${Departamento de F\'isica, ICEx}, {Universidade Federal de Minas Gerais}, {Belo Horizonte}, {MG}, {31270-901}, {Brazil}.}}
\nsfootnote{\textit{$^{b}${Programa de Pós-Graduação em Inovação Tecnológica}, {Universidade Federal de Minas Gerais}, {Belo Horizonte}, {MG}, {31270-901}, {Brazil}.}}
\nsfootnote{\textit{$^{c}${Departamento de Oftalmologia, Faculdade de Medicina, Universidade Federal de Minas Gerais, Belo Horizonte, MG,31270-901, Brazil}.}}
\nsfootnote{\textit{$^{d}${Departamento de Sa\'ude Mental, Faculdade de Medicina, Universidade Federal de Minas Gerais, Belo Horizonte, MG, 31270-901, Brazil}.}}
\nsfootnote{\textit{$^{e}${Departamento de Bioqu\'imica e Imunologia, Universidade Federal de Minas Gerais, Belo Horizonte, MG, 31270-901, Brazil}.}}

\nsfootnote{\dag~\hyperlink{page.11}{Supplementary Information} (SI) avaiable.}
\nsfootnote{\ddag~These authors contributed equally to this work.}
\nsfootnote{$\ast$~E-mail: lmalard@fisica.ufmg.br.}

\begin{abstract} 

\noindent{Given the long subclinical stage of Alzheimer's disease (AD), the study of biomarkers is relevant both for early diagnosis and the fundamental understanding of the pathophysiology of AD. Biomarkers provided by Amyloid-$\upbeta$ (A$\upbeta$) plaques have led to an increasing interest in characterizing this hallmark of AD due to its promising potential. In this work, we characterize A$\upbeta$ plaques by label-free multimodal imaging: we combine two-photon excitation autofluorescence (TPEA), second harmonic generation (SHG), spontaneous Raman scattering (SpRS), coherent anti-Stokes Raman scattering (CARS), and stimulated Raman scattering (SRS) to describe and compare high-resolution images of A$\upbeta$ plaques in brain tissues of an AD mouse model. Comparing single-laser techniques images, we discuss the origin of the SHG, which can be used to locate the plaque core reliably. We study both the core and the halo with vibrational microscopy and compare SpRS and SRS microscopies for different frequencies. We also combine SpRS spectroscopy with SRS microscopy and present two core biomarkers unexplored with SRS microscopy: phenylalanine and amide B. We provide high-resolution SRS images with the spatial distribution of these biomarkers in the plaque and compared them with images of the amide I distribution. The obtained spatial correlation corroborates the feasibility of these biomarkers in the study of A$\upbeta$ plaques. Furthermore, since amide B enables rapid imaging, we discuss its potential as a novel fingerprint for diagnostic applications.}

\end{abstract}

\maketitle


\section*{Introduction}
\noindent Since the first report by Alois Alzheimer \cite{intro1}, Alzheimer’s disease (AD) has been described as a progressive neurodegenerative disorder characterized by cognitive and behavioral impairments, ultimately leading to death \cite{intro1,intro2}. For a long time, its diagnosis relied on the analysis of changes present when cognitive impairment was already installed. Nevertheless, advances in basic research show that AD has a long subclinical stage, with brain changes preceding the onset of symptoms \cite{intro2,intro3}. One of the efforts in the study of AD is the characterization of such disease-related structural changes with the potential to indicate the state of the progression before clinical manifestations \cite{intro2,intro3}. Such effort led to the search for biomarkers through imaging of extracellular amyloid-$\upbeta$ plaques, a pathological hallmark of AD, which can allow a better understanding of the AD pathophysiology and provide new perspectives into its early diagnosis \cite{intro2,intro3,intro4}. 

Amyloid-$\upbeta$ (A$\upbeta$) plaques consist of a core formed predominantly by misfolded A$\upbeta$ peptides, rich in $\upbeta$-sheet conformation, surrounded by a predominantly lipidic halo, in the brain extracellular matrix \cite{intro2,intro3,intro4,intro5}. Due to their structural complexity, several imaging methods have been implemented with the aim to provide a better characterization of A$\upbeta$ plaques. Xenobiotic stain-based approaches have been extensively applied, using exogenous compounds such as methoxy-XO4, Congo red, and Thioflavin S (ThioS) \cite{intro5,intro6,intro7,intro8}. Due to the specificity of the ligands, they provide a limited amount of information. As a result, it is necessary to use different labels to obtain information on different biomolecules, as in the case of using the immunolabels Iba1, GFAP, and Lamp1 to study the halo \cite{shg5,shg6,alz8}. Moreover, the labels can interfere with the information of interest and hamper the interpretation of relevant data \cite{intro9,intro10}, reinforcing the use of label-free imaging techniques.

Nonlinear and vibrational optical imaging have been increasingly used in biology and medicine due to its label-free character, sensitivity, three-dimensional optical sectioning, and high spatial resolution \cite{intro16,intro17}. Furthermore, it has been shown that multimodal imaging, in which several techniques are combined, can be used to maximize the gain of information from complex biological systems \cite{intro11,intro12,intro13,intro14,intro15}. Previous studies of A$\upbeta$ plaques reported label-free imaging by multiphoton fluorescence \cite{auto2009,auto2019}, second harmonic generation (SHG) \cite{shg3}, spontaneous Raman scattering (SpRS) \cite{intro5}, coherent anti-Stokes Raman scattering (CARS) \cite{intro18,intro19}, and, more recently, stimulated Raman scattering (SRS) \cite{intro20}. In some studies of brain imaging obtained from mice, autofluorescence, SHG showing the protein-rich core, and CARS images showing the lipid-rich halo, are presented with limited spatial resolution \cite{auto2009,auto2019,shg3,intro18,intro19}. While the SRS images are presented with higher spatial resolution \cite{intro20}, the investigation concerned the feasibility of narrowband SRS microscopy to study plaques, restricting the core imaging to the amide I vibrational mode. Here, we perform a multimodal optical characterization, exploring the optical properties of endogenous fluorescent biomolecules, molecular structure, and molecular vibrations of A$\upbeta$ plaques to implement high-resolution imaging based on two-photon excitation autofluorescence (TPEA), SHG, SpRS, CARS, and SRS. We provide a two-photon excitation fluorescence (TPEF) image of the ThioS stained tissue as a gold standard procedure to validate our label-free imaging. With a higher spatial resolution, we compare the TPEF ThioS staining, TPEA, and SHG images, which highlight the protein-rich core of the plaques, and also discuss the origin of the SHG signal in A$\upbeta$ plaques. Comparing SpRS and SRS images for various frequencies of different plaques, we observe an almost identical correspondence, even in different scattering geometries. Among those frequencies, we compare the spatial distribution of Phenylalanine (Phe) and amide I in the plaque, showing that Phe, not yet explored in SRS microscopy, is very specific for the core, characterizing a possible biomarker in the low-frequency region in addition to amide I. Combining Raman spectroscopy with SRS microscopy, we look for possible biomarkers in the high-frequency region. We present the plaque image based on the unsaturated lipids vibrational mode, which exhibit a good correlation with other halo vibrations. We propose a core biomarker based on the amide B vibrational mode, also not yet explored in SRS (nor CARS and SpRS) microscopy. We also compare the spatial distribution of amide B and amide I in the plaque. Since amide B enables rapid imaging, this novel fingerprint could be used for diagnostic applications.

\section*{Results and discussion}

\begin{figure*}[tp!]
    \centering
    \includegraphics[width=1\textwidth]{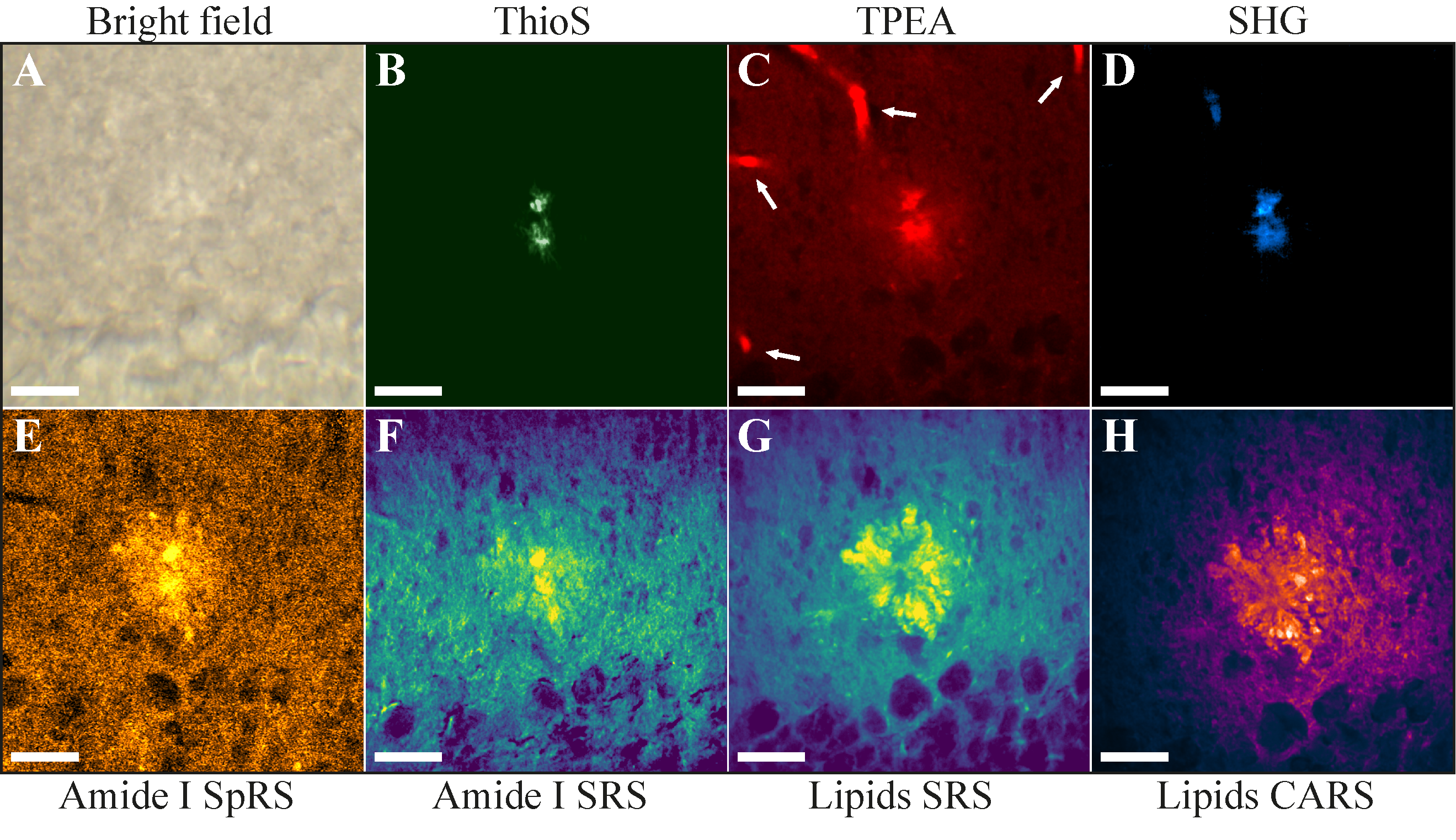}
    \caption{High-resolution multimodal imaging of an A$\upbeta$ plaque from the hippocampus of a 12-month-old mouse brain tissue. (A) Bright field zoomed image from Fig. \ref{figM}\hyperref[figM]{F}. (B) TPEF image of the tissue after ThioS staining. (C) TPEA and (D) SHG images. (E) SpRS image at 1675 cm$^{-1}$. (F) SRS image at 1675 cm$^{-1}$ and (G) 2850 cm$^{-1}$. (H) CARS image at 2850 cm$^{-1}$. All images are of the same region. All scale bars indicate $20$ $\mu$m.}
    \label{fig1}
\end{figure*}

Figure \ref{fig1} shows the zoom at the A$\upbeta$ plaque shown by the blue square in Fig. \ref{figM}\hyperref[figM]{F} (\hyperref[MM]{Materials and Methods}) acquired by using different microscopic modalities. All techniques enable the identification of the A$\upbeta$ plaque, contrary to the bright field image in Fig. \ref{fig1}\hyperref[fig1]{A}, which exhibits no features. A comparison of the findings of multimodal imaging techniques (Figs. \ref{fig1}\hyperref[fig1]{C}-\hyperref[fig1]{H}) with TPEF ThioS staining (Fig. \ref{fig1}\hyperref[fig1]{B}), established as a gold standard, allows an unambiguous plaque identification. 

\subsection*{Two-photon excitation autofluorescence microscopy}
Figure \ref{fig1}\hyperref[fig1]{C} shows the TPEA image obtained with the laser excitation wavelength at $810$ nm and no filters at the detection in order to collect the entire emission window. There is a small contribution of the SHG emission (see below) to the TPEA image, and therefore the SHG is subtracted from the TPEA image, generating Fig. \ref{fig1}\hyperref[fig1]{C}. The origin of TPEA of biological tissues can be attributed to endogenous fluorophores such as nicotinamide adenine dinucleotide, flavin adenine dinucleotide, flavoproteins, tyrosine, tryptophan, collagen, elastin, and lipopigments, to name a few \cite{auto2019, auto2009, auto2006, auto2005, auto1996}. These fluorophores emit light across the visible region with overlapping emissions \cite{auto2019, auto2005}. Lipopigments such as lipofuscin, present in neurons, and associated with age-related neurodegeneration \cite{auto2019, auto2018}, exhibit a broad spectrum, typically in the entire visible range \cite{auto2019, auto2005, auto2018}. For this reason, although the A$\upbeta$ plaque is present in the TPEA image of Fig. \ref{fig1}\hyperref[fig1]{C}, the precise contribution of autofluorescent biomolecules to the observed signal is still under discussion, and proper characterization of biomarkers is a non-trivial task. Moreover, the extraction of information from the TPEA image only by optical characterization can be impractical \cite{auto1996}. Additionally, since those autofluorescent biomolecules are not exclusive to A$\upbeta$ plaques, other bright features are also present. In our TPEA images, blood vessels are observed, highlighted by the arrows in Fig. \ref{fig1}\hyperref[fig1]{C}.

\subsection*{Second harmonic generation microscopy} Another technique that can also be implemented using a single laser wavelength is SHG microscopy. SHG can be present in materials lacking inversion symmetry \cite{boyd,shen}. This signal relies on the anisotropy that particular biological structures exhibit, giving rise to appreciable second-order susceptibility values \cite{shg1,shg13,shg14,shg15}. Fig. \ref{fig1}\hyperref[fig1]{D} shows an SHG image acquired with the laser excitation wavelength at $810$ nm and a bandpass filter centered at $405$ nm ($10$ nm width). In this case, the narrow bandpass filter is important to block the TPEA emission. As shown in Figs. \ref{fig1}\hyperref[fig1]{C}-\hyperref[fig1]{D}, these techniques can be reliable in identifying the A$\upbeta$ core without any extrinsic fluorescent label. Nevertheless, the molecular origin of the SHG and proper characterization of symmetry-based biomarkers provided by A$\upbeta$ plaques is also still under discussion. Biological structures that typically produce SHG include collagen, microtubules, and other non-centrosymmetric proteins \cite{shg1,shg13,shg14,shg15}. In the brain, collagen fibrils have been suggested to affect the detection of amyloid fibrils \cite{shg2}, and also that the origin of SHG in A$\upbeta$ plaques can be attributed to collagen \cite{shg3}, or microtubules in neurites of plaques \cite{auto2019}. However, our image in Fig. \ref{fig1}\hyperref[fig1]{D} clearly shows that the primary source of SHG is the core, where it would not be expected to observe interaction with collagen and where there are no neurites. Previous studies with immunohistochemistry of neurites, astrocytes, and microglia, demonstrated that these cells are present in the surrounding halo, rather than in the core of the A$\upbeta$ plaques \cite{shg5, shg6, alz8, shg7}. Additionally, comparing SHG, TPEA, and TPEF ThioS staining images in Figs. \ref{fig1}\hyperref[fig1]{B}-\hyperref[fig1]{D} allows us to observe similarities in the core images, suggesting a common origin. Furthermore, ThioS is considered a specific label for the core of A$\upbeta$ plaques, being a characteristic ligand of the $\upbeta$-sheet conformation \cite{shg5,shg4}. Studies of $\upbeta$-sheet conformation of \textit{in vitro} and \textit{in situ} amyloid structures suggest that the origin of SHG emission is due to interfacial and symmetry properties of the protein structures \cite{shg8,shg9,shg12}, which is supported by our core image. Therefore, we consider that the origin of the SHG is better associated with these symmetry aspects of protein structures in the core rather than due to the above-mentioned suggestions regarding collagen, neurites, and microtubules. 

In addition to plaque identification based on techniques that highlight the core, such as TPEA and SHG, a full biochemical characterization of the core and halo requires techniques that can retrieve chemical information.

\subsection*{Vibrational microscopy} In vibrational microscopy, the plaque core can be accessed through the C=O stretching mode of amide I \cite{intro5,intro20,amideI1, amideI2}. The misfolding of $A\upbeta$ peptides, one of the leading events for plaque formation, can be identified by the blueshifted ($\sim$ 15 cm$^{-1}$) amide I on SpRS, consistent to a $\upbeta$-sheet secondary structure conformation in relation to the $\alpha$-helix secondary structure peak (at 1660 cm$^{-1}$). Figure \ref{fig1}\hyperref[fig1]{E} shows a SpRS image set at such a vibrational fingerprint (1675 cm$^{-1}$), in which the core is clearly visible, and therefore can be used as a label-free imaging technique. Since SpRS is an incoherent effect, its major disadvantage is that a hyperspectral image like the one shown in Fig. \ref{fig1}\hyperref[fig1]{E} took approximately 20 hours, making this a time-consuming technique for histopathological studies and unfeasible for \textit{in vivo} studies. On the other hand, such a disadvantage is circumvented by coherent nonlinear vibrational microscopy approaches, which keep the rich chemical information provided by Raman spectroscopy without the drawback of low throughput \cite{intro17,intro20}. Figures \ref{fig1}\hyperref[fig1]{F}-\hyperref[fig1]{G} show the SRS images taken at 1675 cm$^{-1}$  (amide I) and 2850 cm$^{-1}$, respectively. The 2850 cm$^{-1}$ Raman peak is ascribed to the C-H stretching modes present in lipid-rich regions, allowing Raman-based microscopies to identify the halo around the core as shown in Fig. \ref{fig1}\hyperref[fig1]{G}. Figure \ref{fig1}\hyperref[fig1]{H} shows the CARS image tuned at 2850 cm$^{-1}$, also revealing the lipid-rich halo around the plaque core. For the CARS image, we have used a bandpass filter (660/13 mn) to measure only the CARS contribution; however, there is a residual TPEA signal from the core. Hence, although at the same vibrational frequency of the SRS (Fig. \ref{fig1}\hyperref[fig1]{G}), the CARS image also shows a signal from the core. Additionally, the implemented homodyne detection scheme does not prevent non-resonant background signals \cite{intro20}. Furthermore, due to the poor sensitivity of our PMT in the near-infrared region, our experimental setup for CARS does not allow us to measure vibrational modes below 2500 cm$^{-1}$. 

\begin{figure*}[tp!]
    \centering
    \includegraphics[width=1\textwidth]{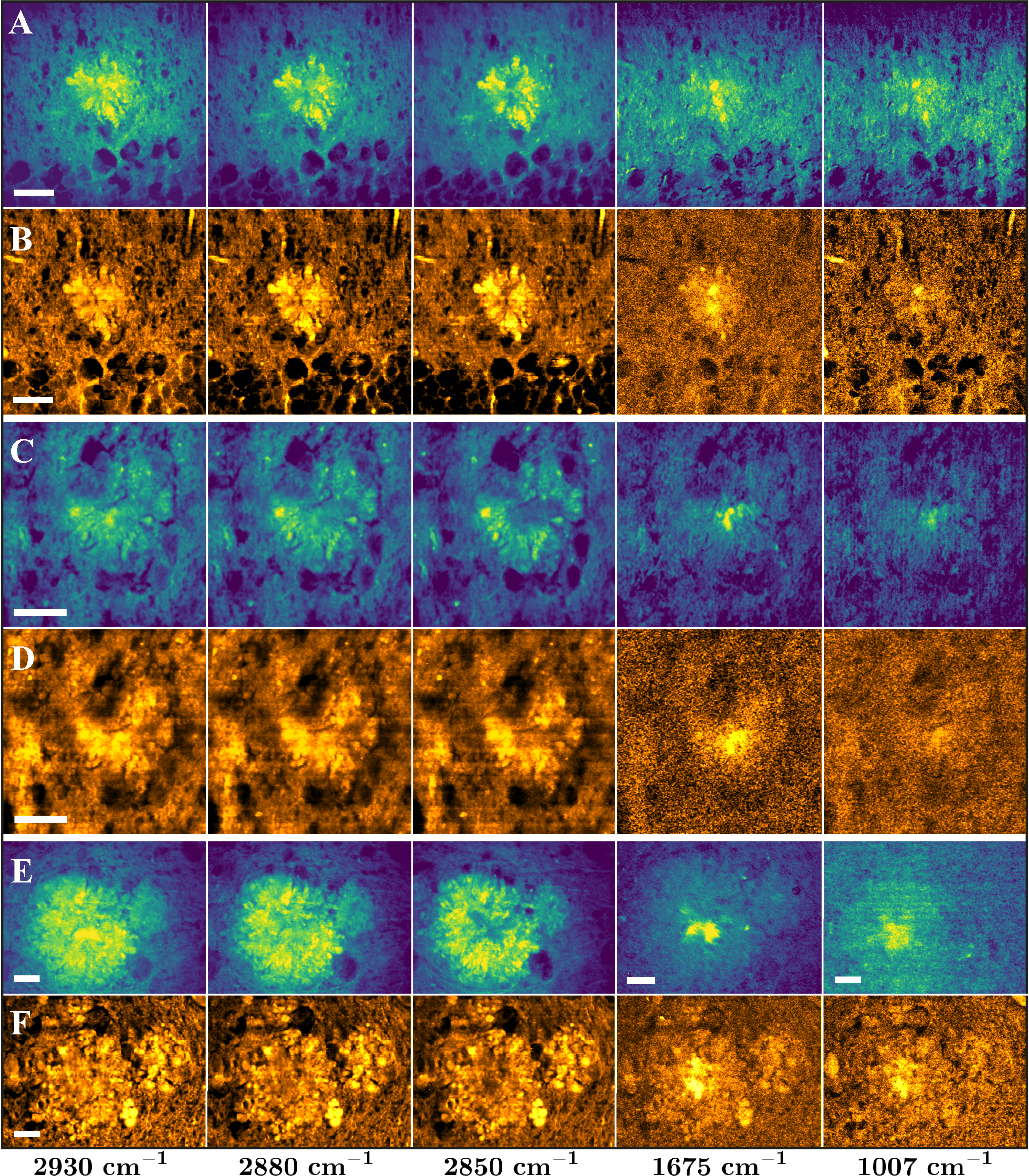}
    \caption{Comparison of narrowband stimulated Raman microscopy and spontaneous Raman microscopy in different scattering geometries. Three different A$\upbeta$ plaques imaged by SRS (A, C, E) and hyperspectral SpRS (B, D, F), taken at different vibrational energies, as listed at the bottom of the images. All scale bars are 20 $\mu$m. Images without a scale bar share the scale bar of the leftmost image in the same line.
}
    \label{fig2}
\end{figure*}

As shown in Fig. \ref{fig1}, label-free, nonlinear optical microscopies such as TPEA and SHG can be very useful to locate A$\upbeta$ plaques; however, it is limited for retrieving chemical information. Vibrational microscopies circumvent this limitation, also allowing to locate other structures, such as the halo shown in Figs. \ref{fig1}\hyperref[fig1]{G}-\hyperref[fig1]{H}. Additionally, since the SRS signal is unaffected by a non-resonant background and proportional to the scatterers concentration, SRS and SpRS share virtually identical spectra \cite{intro17,intro20}. We show in Fig. \ref{fig2} a comparison of SRS and SpRS images for different vibrational frequencies of the plaque in Fig. \ref{fig1} (Figs. \ref{fig2}\hyperref[fig2]{A}-\hyperref[fig2]{B}) and of other two different plaques (Figs. \ref{fig2}\hyperref[fig2]{C}-\hyperref[fig2]{F}). The plaques were located at the hippocampus (Figs. \ref{fig2}\hyperref[fig2]{A}-\hyperref[fig2]{D}) and at the cortex (Figs. \ref{fig2}\hyperref[fig2]{E}-\hyperref[fig2]{F}). We also show a comparison of SRS and SpRS spectra from 2800 to 3075 cm$^{-1}$ (SI, \hyperlink{page.13}{Fig. S2}).

The broad Raman band from 2800 to 3000 cm$^{-1}$ (SI, \hyperlink{page.13}{Fig. S2}) is commonly found in organic samples, as it is the composition of important chemical vibrations, such as the CH$_2$ stretching modes at $\sim$ 2850 cm$^{-1}$ (lipids) and $\sim$ 2880 cm$^{-1}$ (proteins/lipids), and CH$_3$ stretching mode at $\sim$ 2930 cm$^{-1}$ (proteins/lipids) \cite{intro5}. Oppositely to the case of SHG, TPEA, and TPEF ThioS staining, the halo can be imaged clearly by SRS and SpRS: at 2850 cm$^{-1}$ only the halo is visible. By changing to higher frequencies, the halo vibrations start to overlap with vibrational modes of proteins and, therefore, the plaque core is also present in the SRS and SpRS images taken at 2880 cm$^{-1}$ and 2930 cm$^{-1}$, as shown in Fig. \ref{fig2}. These are the most intense Raman modes in the plaques \cite{intro5}, and give rise to signals of lipids forming the halo surrounding the A$\upbeta$ core, whose correlation with dystrophic neurites, microglia and astrocytes has been extensively studied \cite{shg5, shg6, alz8, shg7}. Previous studies have explored the relation between neuroinflammation and AD \cite{alz1,alz2}, which is associated with the presence of astrocytes, microglia, and dystrophic neurites distributed in the halo surrounding the A$\upbeta$ fibrillar deposits \cite{shg5,shg6,alz3,alz5}. The influence of size, location, speed, and extent of plaque growth and their relation to the evidence of neurotoxicity of the plaques was reported in a sequential analysis of the area of neuritic dystrophy in the halo \cite{alz8}. These findings highlight the need for tools capable of characterizing halo biomarkers and show how the specific information provided by SRS and SpRS microscopies can be useful, especially regarding plaque expansion and neurodegeneration progression.

\begin{figure*}[tp!]
    \centering
    \includegraphics[width=1\textwidth]{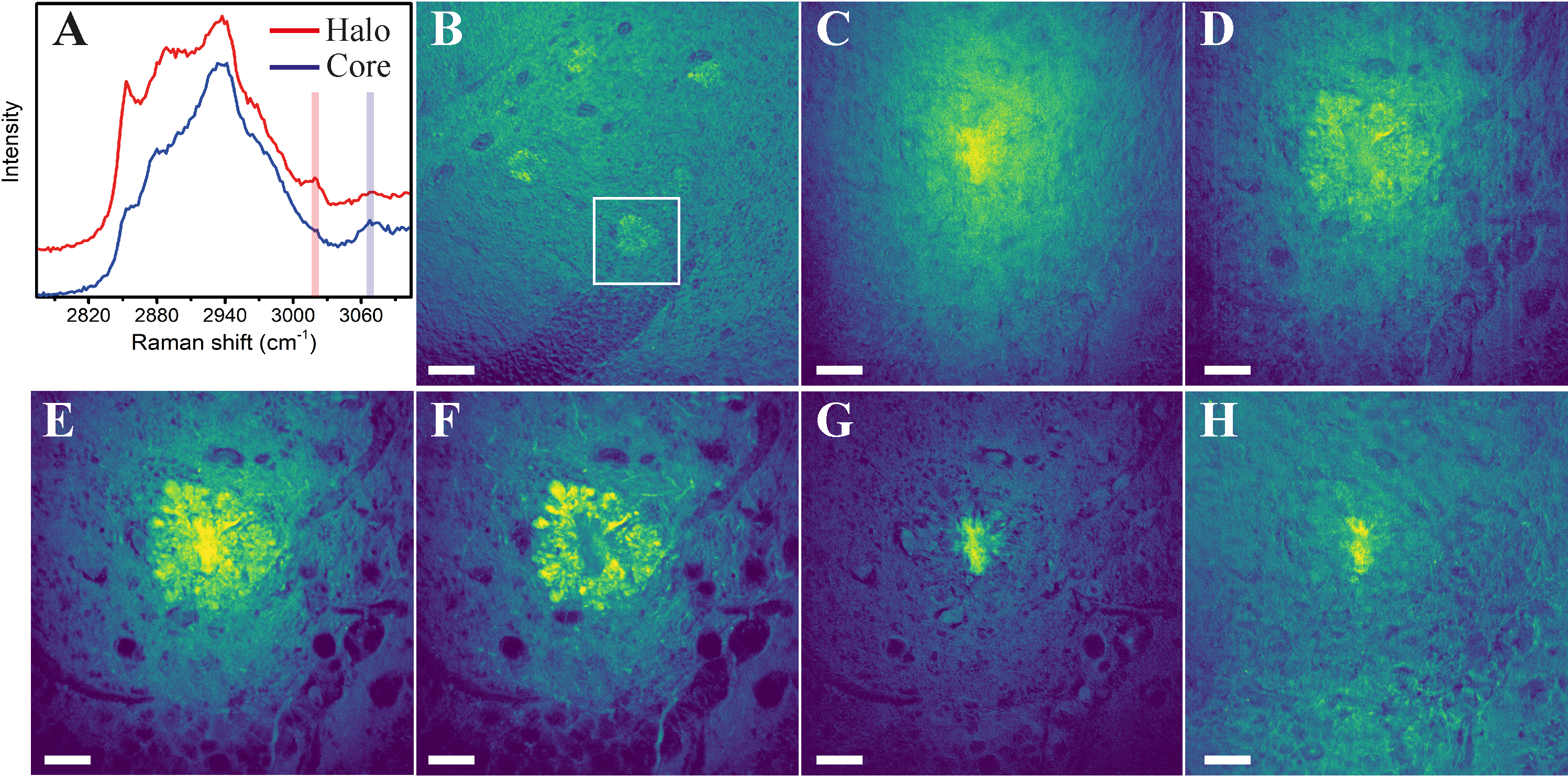}
    \caption{Amide B and unsaturated lipids as core and halo biomarkers, respectively, in the high-frequency region. (A) Typical SpRS spectrum of the halo (upper, red) and core (lower, blue) in the high-frequency region. The purple vertical bar identifies the frequency region of the SRS image in (C), while the orange bar identifies the frequency region of the image in (D). (B) Overview SRS image of the hippocampus with several A$\upbeta$ plaques. The white rectangle identifies the plaque in (C-H). Scale bar is 60 $\mu$m. 
    (C) SRS image taken at 3070 cm$^{-1}$, attributed to amide B vibration. (D) SRS image taken at 3019 cm$^{-1}$, attributed to unsaturated lipids and exhibiting a good correlation with lipids vibration in (F). (E) SRS image taken at 2930 cm$^{-1}$ (protein/lipids), (F) SRS image taken at 2850 cm$^{-1}$ (lipids), and (G) the subtraction of images in (E) and (F). (H) SRS image taken at 1675 cm$^{-1}$, attributed to amide I vibration. All scale bars in (C)-(H) are 20 $\mu$m.
}
    \label{fig3}
\end{figure*}

\subsubsection*{Vibrational biomarkers} SRS and SpRS microscopies also take a step further in the study of the plaque core compared to TPEA and SHG. The overlapping autofluorescent signals from different biomolecules hampers the tracing of the origin of autofluorescence in the plaque only by optical methods, and the origin of the SHG signal is still uncertain. Due to the well-established chemical specificity of SRS and SpRS, it is possible to resolve different signals produced in the core. The fourth column (from left to right) of Fig. \ref{fig2} shows the amide I (1675 cm$^{-1}$) frequency, through which the core is identified. The amide I band is a typical protein band in Raman spectra associated with the polypeptide backbone \cite{amideI1, amideI2}. To date, it has been the only vibrational signature used to identify the core with SRS microscopy \cite{intro20}, which raises the question about the feasibility of using new signatures and the information they can reveal about such a complex structure as A$\upbeta$ plaques.

In this regard, we image the plaque through the vibrational mode of the phenylalanine (Phe, 1007 cm$^{-1}$, last column of Figs. \ref{fig2}\hyperref[fig2]{A}-\hyperref[fig2]{F}), a hydrophobic, aromatic amino acid found in two positions of the A$\upbeta$ peptide primary structure \cite{amideI1,phenyl1}. Its position and contact with other amino acids seem to have a fundamental role in the $\upbeta$-sheet fibril conformation \cite{phenyl1,phenyl2}. Furthermore, a significant difference of Phe concentration was obtained between the core and surrounding tissue \cite{intro5}, making the image of the core clear, which suggests its use in the study of A$\upbeta$ plaques. We also show a comparison between the spatial distribution of Phe and amide I in the plaque (SI, \hyperlink{page.14}{Fig. S3}), showing the specificity of Phe in the core. Therefore, we regard Phe as a core biomarker in the low-frequency region in addition to amide I.

In all our images in Fig. \ref{fig2}, there is a good correspondence between SRS and SpRS microscopies. The slight differences between the two are due to the scattering geometry: the measurement of SRS was performed in a transmission geometry, in contrast to SpRS, measured in a reflection mode. Additionally, the SpRS images for the two core biomarkers (1675 and 1007 cm$^{-1}$) were background subtracted to enhance the contrast. Nevertheless, the SpRS and SRS spectra from 2800 to 3075 cm$^{-1}$ match very well (SI, \hyperlink{page.13}{Fig. S2}). Therefore, all the chemical information delivered by spontaneous Raman spectroscopy can be used for narrowband SRS imaging without loss of information. This demonstrates that we can combine the chemical specificity and full spectral data of SpRS spectroscopy with the speed and resolution of SRS microscopy reliably and straightforwardly.

An important difference between the low-frequency core and high-frequency halo biomarkers presented here is that the latter has the highest intensity. For this reason, most studies with coherent Raman microscopy were carried out in this region \cite{intro20}. To investigate vibrational signatures of the core in the high-frequency region, we further applied the combination of Raman spectroscopy and SRS microscopy. Figure \ref{fig3}\hyperref[fig3]{A} shows typical SpRS spectra from the core (blue) and the halo (red) of an A$\upbeta$ plaque at the high-frequency region. This region is usually unexplored to obtain information about proteins. Nevertheless, besides the already commented Raman bands from 2800 to 3000 cm$^{-1}$, there is a pronounced Raman band between 3050 and 3090 cm$^{-1}$ with a peak at $\sim$ 3070 cm$^{-1}$ in the core spectrum. This signal has been assigned by infrared spectroscopy as the amide B vibrational mode present in proteins due to N-H stretching modes \cite{amideb1}. It is usually part of a Fermi resonance doublet (amide A and B), and it is resonant with an amide II combination mode in $\upbeta$-sheet conformation \cite{amideI2,amideb1}. This is of substantial significance since the amide II band is considered too weak or absent in the Raman spectrum \cite{amideI2,amideb1}, but it provides valuable structural information, and it is suggested for secondary structure prediction as a counterpart of amide I \cite{amideb1,amideb2}. Figure \ref{fig3}\hyperref[fig3]{B} shows an overview SRS image of a hippocampus with several A$\upbeta$ plaques based on lipids CH$_2$ stretching mode (2850 cm$^{-1}$) and acquired with a 20X objective (numerical aperture, 0.75). The white rectangle identifies the plaque shown in Figs. \ref{fig3}\hyperref[fig3]{C}-\hyperref[fig3]{H}. Figure \ref{fig3}\hyperref[fig3]{C} shows the SRS image taken at 3070 cm$^{-1}$, which images the core clearly, thus demonstrating the feasibility of using the amide B frequency to locate the core of the A$\upbeta$ plaques. To the best of our knowledge, no core signature in this high-frequency region has been reported. Such a specific image in this region could only be presented by subtracting the images based on frequencies such as 2930 cm$^{-1}$ (Fig. \ref{fig3}\hyperref[fig3]{E}, proteins/lipids CH$_3$ stretching mode) and 2850 cm$^{-1}$ (Fig. \ref{fig3}\hyperref[fig3]{F}, lipids CH$_2$ stretching mode). Comparing this subtraction in Fig. \ref{fig3}\hyperref[fig3]{G} with the image based on amide I ($\sim$ 1675 cm$^{-1}$) in Fig. \ref{fig3}\hyperref[fig3]{H}, a good correlation is observed. Such a correlation is also present when comparing with amide B image (Fig. \ref{fig3}\hyperref[fig3]{C}), demonstrating the applicability of this vibrational mode. We compare additional plaques images (SI, \hyperlink{page.15}{Fig. S4}), showing the robustness of the method. A comparison of the spatial distribution of amide B and amide I in the plaque is also provided (SI, \hyperlink{page.16}{Fig. S5}). Therefore, we regard amide B as a core biomarker in the high-frequency region.

The importance of these results for microscopy can also be appreciated in two ways: the analysis of different signatures for the core, such as amide I, Phe, and amide B vibrational modes, can reveal complementary information about the plaque, which can be valuable for its basic understanding and its precise role in AD-related neurodegeneration. Also, while amide B is described as a weakly absorbing component in infrared spectroscopy \cite{amideI2,amideb1}, it allowed the acquisition of core images with high contrast, high intensity, and few accumulations with SRS. In our experiment, we obtained SRS images at the amide B frequency with approximately 7 times higher mean contrast than those obtained at the amide I frequency. This result suggests amide B as a novel fingerprint to be incorporated in the study of A$\upbeta$ plaques for diagnostic applications, where the acquisition time is of great relevance. 

Correspondingly, the halo spectrum shows a band between 3010 cm$^{-1}$ and 3025 cm$^{-1}$ (peak at 3019 cm$^{-1}$), which is absent in the core spectrum shown in Fig. \ref{fig3}\hyperref[fig3]{A}. This band is associated with unsaturated $=$CH bindings and is attributed to unsaturated lipids \cite{unslip}. Figure \ref{fig3}\hyperref[fig3]{G} shows the SRS image at 3019 cm$^{-1}$ , which correlates with the characteristic SRS image of the halo at 2850 cm$^{-1}$ in Fig. \ref{fig3}\hyperref[fig3]{F}. Therefore, it is also possible to use this Raman band to locate the halo by SRS imaging in the high-frequency region. We also provide images based on the same vibrations shown in Figs. \ref{fig3}\hyperref[fig3]{C}-\hyperref[fig3]{H} for two other plaques in the hippocampus (SI, \hyperlink{page.17}{Fig. S6}).

\section*{Conclusion}

In summary, we performed label-free multimodal imaging by nonlinear and vibrational microscopies to study and present high-resolution images of A$\upbeta$ plaques in the hippocampus and cortex of brain tissues of bitransgenic mice AD model. We validated the presence of A$\upbeta$ plaques by TPEF ThioS staining. While imaging with exogenous labels is undesirable, autofluorescence may be non-ideal in some applications due to the different overlapping signals. We presented SHG imaging of the core and compared with TPEF ThioS staining and TPEA images, which, together with an analysis of results in the literature, suggest a protein-related origin of the SHG signal. Also, using vibrational microscopies based on SpRS, CARS, and SRS, we studied both the core and the halo of the plaques. While the CARS imaging is affected by spurious backgrounds, we showed that SpRS and SRS microscopies produce virtually identical images, even in different experimental setups. This allows for a straightforward combination of Raman spectroscopy information with the advantages of SRS microscopy. We also presented a halo biomarker based on unsaturated lipids and two core biomarkers not yet reported in SRS microscopy studies of A$\upbeta$ plaques: phenylalanine (Phe), in the low-frequency region, and amide B, in the high-frequency region. We compared the spatial distribution of the three core biomarkers. While the Phe image appears to be more specific to the central part of the core than amide I, amide B allowed us to obtain images with higher acquisition rates than amide I and Phe. This result suggests amide B as a novel fingerprint in the study of A$\upbeta$ plaques for diagnostic applications.

\section*{Materials and methods}\label{MM}

\begin{figure*}[tp!]
    \centering
    \includegraphics[width=1\textwidth]{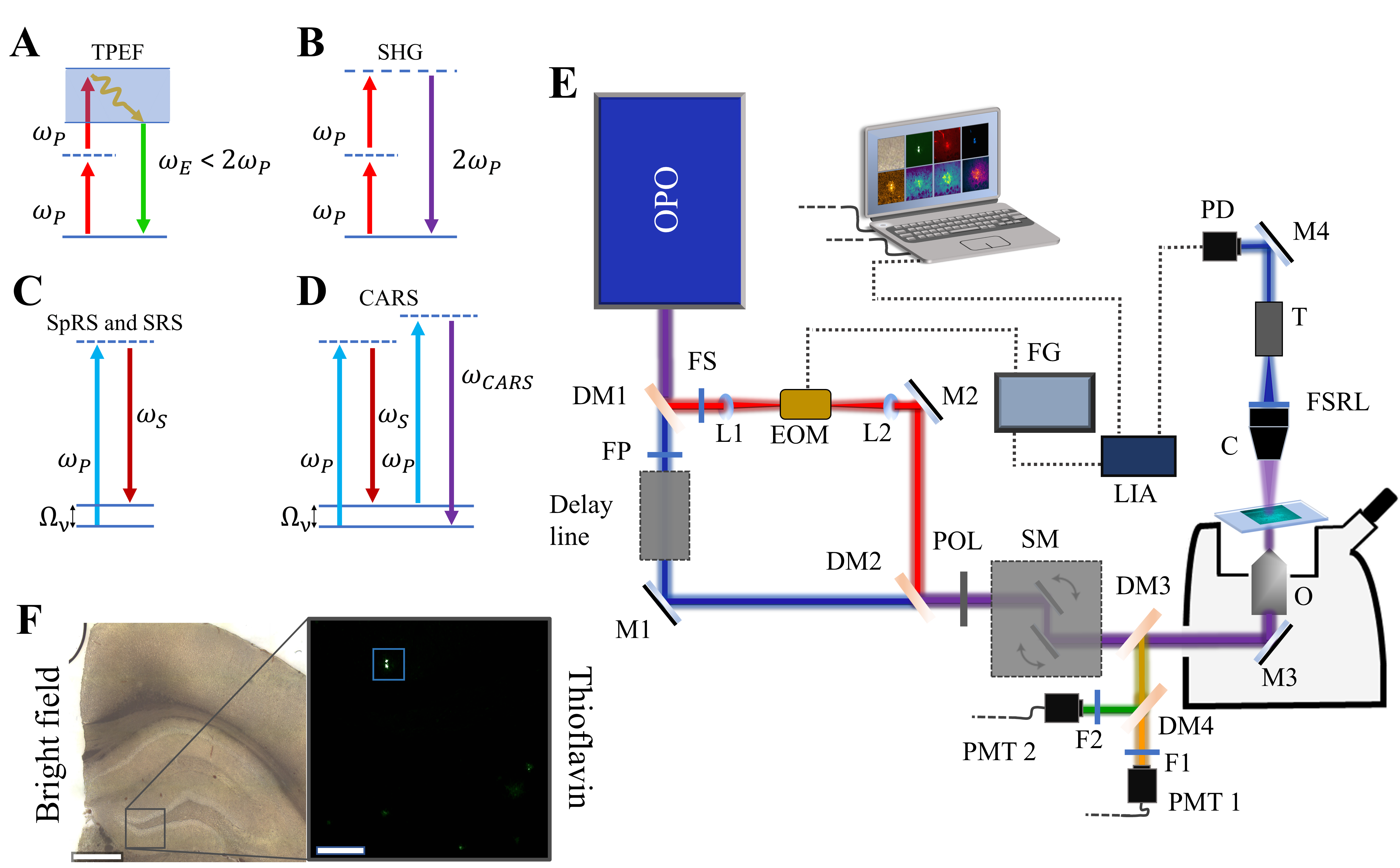}
    \caption{Optical effects, experimental implementation, and Thioflavin validation. Energy level description for (A) TPEF, (B) SHG, (c) SpRS and SRS, and (D) CARS. (E) Experimental setup. (F) Left side: bright field image of the mouse brain slice (scale bar is 500 $\mu$m). Right side: A$\upbeta$ plaque highlighted in the TPEF ThioS staining image of the hippocampus region shown by the gray square in the bright field image (scale bar is 100 $\mu$m).}
    \label{figM}
\end{figure*}

\subsection*{Animal model and tissue preparation}

APP:PS1 double transgenic male mice (Tg) were purchased from the Jackson Laboratories. These mice have the human amyloid precursor protein gene containing the Swedish mutation K594N/M595L and delta E9 mutant human presenilin 1 gene, causing accelerated A$\upbeta$ deposition in the brain after 5 months of age \cite{prep1997, prep2001, prep2006}. For this study, we used mice of 6 and 12 months old. The mice were anesthetized and sacrificed, then had the brain removed and kept overnight in 4\% paraformaldehyde (PFA), followed by cryosectioning of the frontal cortex and hippocampus (40 $\mu$m thick slices). These slices were washed three times with phosphate-buffered saline (PBS) and covered with a coverslip to spectroscopic analyses. After the multimodal imaging, tissue sections were stained using 1\% ThioS solution (w/v) in 50\% ethanol (v/v) (Sigma-Aldrich). The samples were immersed for 8 minutes in ThioS solution, immediately immersed three times in 50\% ethanol, washed with PBS, and seeded directly onto coverslips for TPEF imaging. Figure \ref{figM}\hyperref[figM]{F} shows a TPEF ThioS staining image of a sample obtained from the hippocampus containing several plaques. All images in Fig. \ref{fig1} refer to the one highlighted by the blue square in Fig. \ref{figM}\hyperref[figM]{F}. Our investigation is in agreement with the Guide for the Care and Use of Laboratory Animals and was approved by the Ethics Committee for Animal Utilization in Research (CEUA) of the Federal University of Minas Gerais (protocol 225/2014), under the criteria of the National Animal Experimentation Control Council (CONCEA). 

\subsection*{Optical effects and multimodal plataform}

Figures \ref{figM}\hyperref[figM]{A}-\hyperref[figM]{D} shows the exchange of energy diagrams for the optical effects studied in this work. Fig. \ref{figM}\hyperref[figM]{A} shows the TPEF, where a high-energy electronic state becomes accessible through the simultaneous absorption of two low-energy photons at frequency $\omega_p$. In this third-order nonlinear effect, part of the absorbed energy is lost in non-radiative processes, while the other part is converted into a spontaneous fluorescent emission at frequency $\omega_E$. Due to the non-radiative processes, such fluorescent emission occurs at a frequency lower than the sum of the absorbed frequencies\cite{boyd,shen}. Figure \ref{figM}\hyperref[figM]{B} shows the SHG, a second-order nonlinear optical effect where the two low-energy photons at $\omega_p$ can be up-converted into one photon with exactly twice the incident frequency. This spontaneous parametric up-conversion can only occur in materials lacking inversion symmetry in the electric dipole approximation \cite{boyd,shen}. Figure \ref{figM}\hyperref[figM]{C} shows the energy diagram for two different, but related effects. When one laser beam of frequency $\omega_p$ interacts with a material, it can undergo a spontaneous inelastic scattering at the frequency $\omega_S$. In the case where $\omega_s < \omega_p$, the material is left in an excited vibrational state at the frequency $\Omega_{\nu}$, which characterizes the Stokes process of SpRS \cite{boyd,shen}. When two laser beams at frequencies $\omega_p$ and $\omega_S$ interact coherently with the material, additional third-order nonlinear optical effects arise. If the beating frequency ($\omega_p - \omega_S$) is resonant with the vibrational transition at $\Omega_{\nu}$, the beam at $\omega_S$ stimulates the conversion of the beam at $\omega_p$ into $\omega_S$, also at the cost of leaving the material in an excited vibrational state. Such a process is the SRS \cite{boyd,shen}. Simultaneously with the SRS, the second beam at $\omega_S$ also stimulates a four-wave mixing process that generates the $\omega_ {aS}$ frequency. In this case, the nonlinear generation of $\omega_ {aS}$ is orders of magnitude more likely to occur than the linear generation of the anti-Stokes Raman scattering, which characterizes the CARS in the diagram of Fig. \ref{figM}\hyperref[figM]{D} \cite{boyd,shen}. All Raman-mediated effects probe vibrational states, therefore providing specific chemical information about the material.

We obtained SpRS images and spectra using the WITec alpha300 SAR confocal system operating with a 532 nm CW laser excitation, focused by a 60X oil objective (numerical aperture, 1.4), and scanning with steps of $\sim$260 nm. We generated both images and spectra with the Project 5 WITec software. All nonlinear optical effects were studied through the setup shown in Fig. \ref{figM}\hyperref[figM]{E}. For TPEF and SHG, we used a pump beam of 180 fs pulse-width and 80 MHz repetition rate, tuned at 810 nm. The beam is transmitted through the dichroic mirrors DM1 and DM2 (following the path of the purple beam) and is focused by a 60X apochromatic oil objective O (numerical aperture, 1.4). The sample image is done through a scanning laser microscope (LaVision Biotec). The dichroic mirrors DM3 and DM4 direct the reflected signal generated by the material to the photomultipliers PMT1 and PMT2. Bandpass filters F1 and F2 can be inserted before the PMT1 and PMT2, respectively, allowing an unambiguous measurement of the effect. For the implementation of SRS and CARS, the optical parametric oscillator OPO (APE picoEMERALD) provides two beams: a 5-6 ps pump beam of tunable frequency $\omega_P$ and a 7 ps Stokes beam at a fixed frequency $\omega_S$, in blue and red, respectively, as in Fig. \ref{figM}\hyperref[figM]{E}. Frequency $\omega_P$ is tuned so that the beat frequency is resonant with the frequency of the vibrational mode. The dichroic mirror DM1 splits the beams in a Mach-Zehnder configuration where the filters FP and FS ensure no contamination of the pump beam (Stokes) in the arm of the Stokes beam (pump), respectively. The electro-optical modulator EOM connected to the function generator FG produces a high-frequency (10 MHz) polarization-modulation in the Stokes beam. The dichroic mirror DM2 recombines both beams in a collinear configuration. Temporal synchronization is achieved through a delay line. The polarizer POL after the recombination of the beams transforms the EOM polarization-modulation into amplitude-modulation and ensures that both the unmodulated (pump) and modulated (Stokes) beams have the same polarization state. The scanning mirrors SM allow the beams to map the sample when focused by the 60X apochromatic oil objective O (numerical aperture, 1.4) in the scanning laser microscope. The backscattered CARS signal is then directed to the PMT1 photomultiplier by the dichroic mirrors DM3 and DM4. A bandpass filter eliminates signal contamination by TPEF. The forward SRS signal is collected by the collimator C and the filter FSRL eliminates the Stokes component of the beam. The telescope T reduces the beam movement on the detector area of the photodiode PD, where the SRS signal is extracted with the help of the lock-in amplifier LIA coupled to the PD and connected to the EOM via the synchronization port of the FG. This setup allows the SRS signal to reach a high signal/noise ratio, as shown by previous works \cite{intro17,intro20}.

After the measurements, we compare our findings with a TPEF image of the tissue stained with ThioS to validate the label-free techniques. This fluorescent label is widely recognized as a dye in neurodegenerative diseases studies, particularly for visualizing amyloid plaques in AD models. \cite{intro5,intro6,intro8,shg5,shg6,alz8}. Figure \ref{figM}\hyperref[figM]{F} shows the bright field image of a section of the brain and the TPEF image of an enlarged section in the hippocampus. The blue square identifies the plaque studied in our multimodal analysis of Fig. \ref{fig1}. A saturated image is also provided (SI, \hyperlink{page.12}{Fig. S1}), favoring the identification of the tissue and the other A$\upbeta$ plaques.

\section*{Author contributions} 
R.C., L.L., and E.A.F. performed the experiments and analyzed the data. R.C., L.L., E.A.F., A.B, A.J., and L.M.M. designed the experiments. M.R.S. provided the early conceptual idea and contributed to the establishment of the colony of mice. R.C. and L.M.M wrote the manuscript. All the authors discussed the results and reviewed the manuscript.

\section*{Conflicts of interest}
The authors declare no competing interest.

\section*{Acknowledgements}
We acknowledge financial support from FINEP (01.13.0330.00), CAPES, Fapemig (TEC - RED-00282-16, APQ-03052-15), CNPq and CNPq project 302775/2018-8.





\makeatletter 
\renewcommand\@biblabel[1]{#1} 
\makeatother
\bibliography{main}

\providecommand*{\mcitethebibliography}{\thebibliography}
\csname @ifundefined\endcsname{endmcitethebibliography}
{\let\endmcitethebibliography\endthebibliography}{}
\begin{mcitethebibliography}{56}
\providecommand*{\natexlab}[1]{#1}
\providecommand*{\mciteSetBstSublistMode}[1]{}
\providecommand*{\mciteSetBstMaxWidthForm}[2]{}
\providecommand*{\mciteBstWouldAddEndPuncttrue}
  {\def\EndOfBibitem{\unskip.}}
\providecommand*{\mciteBstWouldAddEndPunctfalse}
  {\let\EndOfBibitem\relax}
\providecommand*{\mciteSetBstMidEndSepPunct}[3]{}
\providecommand*{\mciteSetBstSublistLabelBeginEnd}[3]{}
\providecommand*{\EndOfBibitem}{}
\mciteSetBstSublistMode{f}
\mciteSetBstMaxWidthForm{subitem}
{(\emph{\alph{mcitesubitemcount}})}
\mciteSetBstSublistLabelBeginEnd{\mcitemaxwidthsubitemform\space}
{\relax}{\relax}

\bibitem[Strassnig and Ganguli(2005)]{intro1}
M.~Strassnig and M.~Ganguli, \emph{Psychiatry (Edgmont)}, 2005, \textbf{2},
  30--33\relax
\mciteBstWouldAddEndPuncttrue
\mciteSetBstMidEndSepPunct{\mcitedefaultmidpunct}
{\mcitedefaultendpunct}{\mcitedefaultseppunct}\relax
\EndOfBibitem
\bibitem[Jellinger(2006)]{intro2}
K.~A. Jellinger, \emph{Journal of Neural Transmission}, 2006, \textbf{113},
  1603–1623\relax
\mciteBstWouldAddEndPuncttrue
\mciteSetBstMidEndSepPunct{\mcitedefaultmidpunct}
{\mcitedefaultendpunct}{\mcitedefaultseppunct}\relax
\EndOfBibitem
\bibitem[Alzheimer's(2018)]{intro3}
A.~Alzheimer's, \emph{Alzheimer's \& Dementia}, 2018, \textbf{14},
  367--429\relax
\mciteBstWouldAddEndPuncttrue
\mciteSetBstMidEndSepPunct{\mcitedefaultmidpunct}
{\mcitedefaultendpunct}{\mcitedefaultseppunct}\relax
\EndOfBibitem
\bibitem[Citron(2002)]{intro4}
M.~Citron, \emph{Nature Neuroscience}, 2002, \textbf{5}, 1055–1057\relax
\mciteBstWouldAddEndPuncttrue
\mciteSetBstMidEndSepPunct{\mcitedefaultmidpunct}
{\mcitedefaultendpunct}{\mcitedefaultseppunct}\relax
\EndOfBibitem
\bibitem[Fonseca \emph{et~al.}(2019)Fonseca, Lafetá, Cunha, Miranda, Campos,
  Medeiros, Romano-Silva, Silva, Barbosa, Vieira, Malard, and Jorio]{intro5}
E.~A. Fonseca, L.~Lafetá, R.~Cunha, H.~Miranda, J.~Campos, H.~G. Medeiros,
  M.~A. Romano-Silva, R.~A. Silva, A.~S. Barbosa, R.~P. Vieira, L.~M. Malard
  and A.~Jorio, \emph{Analyst}, 2019, \textbf{144}, 7049--7056\relax
\mciteBstWouldAddEndPuncttrue
\mciteSetBstMidEndSepPunct{\mcitedefaultmidpunct}
{\mcitedefaultendpunct}{\mcitedefaultseppunct}\relax
\EndOfBibitem
\bibitem[Klunk \emph{et~al.}(2002)Klunk, Bacskai, Mathis, Kajdasz, McLellan,
  Frosch, Debnath, Holt, Wang, and Hyman]{intro6}
W.~E. Klunk, B.~J. Bacskai, C.~A. Mathis, S.~T. Kajdasz, M.~E. McLellan, M.~P.
  Frosch, M.~L. Debnath, D.~P. Holt, Y.~Wang and B.~T. Hyman, \emph{Journal of
  Neuropathology \& Experimental Neurology}, 2002, \textbf{61}, 797--805\relax
\mciteBstWouldAddEndPuncttrue
\mciteSetBstMidEndSepPunct{\mcitedefaultmidpunct}
{\mcitedefaultendpunct}{\mcitedefaultseppunct}\relax
\EndOfBibitem
\bibitem[Wilcock \emph{et~al.}(2006)Wilcock, Gordon, and Morgan]{intro7}
D.~Wilcock, M.~Gordon and D.~Morgan, \emph{Nature Protocols}, 2006, \textbf{1},
  1591–1595\relax
\mciteBstWouldAddEndPuncttrue
\mciteSetBstMidEndSepPunct{\mcitedefaultmidpunct}
{\mcitedefaultendpunct}{\mcitedefaultseppunct}\relax
\EndOfBibitem
\bibitem[McLellan \emph{et~al.}(2003)McLellan, Kajdasz, Hyman, and
  Bacskai]{intro8}
M.~E. McLellan, S.~T. Kajdasz, B.~T. Hyman and B.~J. Bacskai, \emph{Journal of
  Neuroscience}, 2003, \textbf{23}, 2212--2217\relax
\mciteBstWouldAddEndPuncttrue
\mciteSetBstMidEndSepPunct{\mcitedefaultmidpunct}
{\mcitedefaultendpunct}{\mcitedefaultseppunct}\relax
\EndOfBibitem
\bibitem[Condello \emph{et~al.}(2015)Condello, Yuan, Schain, and
  Grutzendler]{shg5}
C.~Condello, P.~Yuan, A.~Schain and J.~Grutzendler, \emph{Nature
  Communication}, 2015, \textbf{6}, 6176\relax
\mciteBstWouldAddEndPuncttrue
\mciteSetBstMidEndSepPunct{\mcitedefaultmidpunct}
{\mcitedefaultendpunct}{\mcitedefaultseppunct}\relax
\EndOfBibitem
\bibitem[Yuan \emph{et~al.}(2016)Yuan, Condello, Keene, Wang, Bird, Paul, Luo,
  Colonna, Baddeley, and Grutzendler]{shg6}
P.~Yuan, C.~Condello, C.~D. Keene, Y.~Wang, T.~D. Bird, S.~M. Paul, W.~Luo,
  M.~Colonna, D.~Baddeley and J.~Grutzendler, \emph{Neuron}, 2016, \textbf{90},
  724 -- 739\relax
\mciteBstWouldAddEndPuncttrue
\mciteSetBstMidEndSepPunct{\mcitedefaultmidpunct}
{\mcitedefaultendpunct}{\mcitedefaultseppunct}\relax
\EndOfBibitem
\bibitem[Condello \emph{et~al.}(2011)Condello, Schain, and Grutzendler]{alz8}
C.~Condello, A.~Schain and J.~Grutzendler, \emph{Scientific Reports}, 2011,
  \textbf{1}, 19\relax
\mciteBstWouldAddEndPuncttrue
\mciteSetBstMidEndSepPunct{\mcitedefaultmidpunct}
{\mcitedefaultendpunct}{\mcitedefaultseppunct}\relax
\EndOfBibitem
\bibitem[Liebscher and Meyer-Luehmann(2012)]{intro9}
S.~Liebscher and M.~Meyer-Luehmann, \emph{Frontiers in Psychiatry}, 2012,
  \textbf{3}, 26\relax
\mciteBstWouldAddEndPuncttrue
\mciteSetBstMidEndSepPunct{\mcitedefaultmidpunct}
{\mcitedefaultendpunct}{\mcitedefaultseppunct}\relax
\EndOfBibitem
\bibitem[Cohen \emph{et~al.}(2009)Cohen, Ikonomovic, Abrahamson, Paljug,
  DeKosky, Lefterov, Koldamova, Shao, Debnath, Mason, and Klunk]{intro10}
A.~D. Cohen, M.~D. Ikonomovic, E.~E. Abrahamson, W.~R. Paljug, S.~T. DeKosky,
  I.~M. Lefterov, R.~P. Koldamova, L.~Shao, M.~L. Debnath, C.~A. Mason, N.
  S.~Mathis and W.~E. Klunk, \emph{Letters in Drug Design \& Discovery}, 2009,
  \textbf{6}, 437--444\relax
\mciteBstWouldAddEndPuncttrue
\mciteSetBstMidEndSepPunct{\mcitedefaultmidpunct}
{\mcitedefaultendpunct}{\mcitedefaultseppunct}\relax
\EndOfBibitem
\bibitem[Zipfel \emph{et~al.}(2003)Zipfel, Williams, and Webb]{intro16}
W.~Zipfel, R.~Williams and W.~Webb, \emph{Nature Biotechnology}, 2003,
  \textbf{21}, 1369–1377\relax
\mciteBstWouldAddEndPuncttrue
\mciteSetBstMidEndSepPunct{\mcitedefaultmidpunct}
{\mcitedefaultendpunct}{\mcitedefaultseppunct}\relax
\EndOfBibitem
\bibitem[Cheng and Xie(2015)]{intro17}
J.-X. Cheng and X.~S. Xie, \emph{Science}, 2015, \textbf{350}, aaa8870\relax
\mciteBstWouldAddEndPuncttrue
\mciteSetBstMidEndSepPunct{\mcitedefaultmidpunct}
{\mcitedefaultendpunct}{\mcitedefaultseppunct}\relax
\EndOfBibitem
\bibitem[Meyer \emph{et~al.}(2011)Meyer, Bergner, Krafft, Akimov, Dietzek,
  Popp, Bielecki, Romeike, Reichart, and Kalff]{intro11}
T.~Meyer, N.~Bergner, C.~Krafft, D.~Akimov, B.~Dietzek, J.~Popp, C.~Bielecki,
  B.~F.~M. Romeike, R.~Reichart and R.~Kalff, \emph{Journal of Biomedical
  Optics}, 2011, \textbf{16}, 1--10\relax
\mciteBstWouldAddEndPuncttrue
\mciteSetBstMidEndSepPunct{\mcitedefaultmidpunct}
{\mcitedefaultendpunct}{\mcitedefaultseppunct}\relax
\EndOfBibitem
\bibitem[Pfeffer \emph{et~al.}(2008)Pfeffer, Olsen, Ganikhanov, and
  Légaré]{intro12}
C.~P. Pfeffer, B.~R. Olsen, F.~Ganikhanov and F.~Légaré, \emph{Journal of
  Structural Biology}, 2008, \textbf{164}, 140 -- 145\relax
\mciteBstWouldAddEndPuncttrue
\mciteSetBstMidEndSepPunct{\mcitedefaultmidpunct}
{\mcitedefaultendpunct}{\mcitedefaultseppunct}\relax
\EndOfBibitem
\bibitem[Heuke \emph{et~al.}(2013)Heuke, Vogler, Meyer, Akimov, Kluschke,
  Röwert-Huber, Lademann, Dietzek, and Popp]{intro13}
S.~Heuke, N.~Vogler, T.~Meyer, D.~Akimov, F.~Kluschke, H.-J. Röwert-Huber,
  J.~Lademann, B.~Dietzek and J.~Popp, \emph{British Journal of Dermatology},
  2013, \textbf{169}, 794--803\relax
\mciteBstWouldAddEndPuncttrue
\mciteSetBstMidEndSepPunct{\mcitedefaultmidpunct}
{\mcitedefaultendpunct}{\mcitedefaultseppunct}\relax
\EndOfBibitem
\bibitem[Mazumder \emph{et~al.}(2019)Mazumder, Balla, Zhuo, Kistenev, Kumar,
  Kao, Brasselet, Nikolaev, and Krivova]{intro14}
N.~Mazumder, N.~K. Balla, G.-Y. Zhuo, Y.~V. Kistenev, R.~Kumar, F.-J. Kao,
  S.~Brasselet, V.~V. Nikolaev and N.~A. Krivova, \emph{Frontiers in Physics},
  2019, \textbf{7}, 170\relax
\mciteBstWouldAddEndPuncttrue
\mciteSetBstMidEndSepPunct{\mcitedefaultmidpunct}
{\mcitedefaultendpunct}{\mcitedefaultseppunct}\relax
\EndOfBibitem
\bibitem[Chen \emph{et~al.}(2009)Chen, Wang, Slipchenko, Jung, Shi, Zhu,
  Buhman, and Cheng]{intro15}
H.~Chen, H.~Wang, M.~N. Slipchenko, Y.~Jung, Y.~Shi, J.~Zhu, K.~K. Buhman and
  J.-X. Cheng, \emph{Optics Express}, 2009, \textbf{17}, 1282--1290\relax
\mciteBstWouldAddEndPuncttrue
\mciteSetBstMidEndSepPunct{\mcitedefaultmidpunct}
{\mcitedefaultendpunct}{\mcitedefaultseppunct}\relax
\EndOfBibitem
\bibitem[Kwan \emph{et~al.}(2009)Kwan, Duff, Gouras, and Webb]{auto2009}
A.~C. Kwan, K.~Duff, G.~K. Gouras and W.~W. Webb, \emph{Optics Express}, 2009,
  \textbf{17}, 3679--3689\relax
\mciteBstWouldAddEndPuncttrue
\mciteSetBstMidEndSepPunct{\mcitedefaultmidpunct}
{\mcitedefaultendpunct}{\mcitedefaultseppunct}\relax
\EndOfBibitem
\bibitem[Wang \emph{et~al.}(2019)Wang, Lin, Lin, Sun, Lin, Huang, Tao, Wang,
  Wu, Chen, and Chen]{auto2019}
S.~Wang, B.~Lin, G.~Lin, C.~Sun, R.~Lin, J.~Huang, J.~Tao, X.~Wang, Y.~Wu,
  L.~Chen and J.~Chen, \emph{Neurophotonics}, 2019, \textbf{6}, 1--11\relax
\mciteBstWouldAddEndPuncttrue
\mciteSetBstMidEndSepPunct{\mcitedefaultmidpunct}
{\mcitedefaultendpunct}{\mcitedefaultseppunct}\relax
\EndOfBibitem
\bibitem[Chakraborty \emph{et~al.}(2020)Chakraborty, Chen, Hsiao, Chiu, and
  Sun]{shg3}
S.~Chakraborty, S.-T. Chen, Y.-T. Hsiao, M.-J. Chiu and C.-K. Sun,
  \emph{Biomedical Optics Express}, 2020, \textbf{11}, 571--585\relax
\mciteBstWouldAddEndPuncttrue
\mciteSetBstMidEndSepPunct{\mcitedefaultmidpunct}
{\mcitedefaultendpunct}{\mcitedefaultseppunct}\relax
\EndOfBibitem
\bibitem[Kiskis \emph{et~al.}(2015)Kiskis, Fink, Nyberg, Thyr, Li, and
  Enejder]{intro18}
J.~Kiskis, H.~Fink, L.~Nyberg, J.~Thyr, J.-Y. Li and A.~Enejder,
  \emph{Scientific Reports}, 2015, \textbf{5}, 13489\relax
\mciteBstWouldAddEndPuncttrue
\mciteSetBstMidEndSepPunct{\mcitedefaultmidpunct}
{\mcitedefaultendpunct}{\mcitedefaultseppunct}\relax
\EndOfBibitem
\bibitem[Lee \emph{et~al.}(2015)Lee, Kim, Song, Oh, and Ko]{intro19}
J.~H. Lee, D.~H. Kim, W.~K. Song, M.-K. Oh and D.-K. Ko, \emph{Journal of
  Biomedical Optics}, 2015, \textbf{20}, 1--7\relax
\mciteBstWouldAddEndPuncttrue
\mciteSetBstMidEndSepPunct{\mcitedefaultmidpunct}
{\mcitedefaultendpunct}{\mcitedefaultseppunct}\relax
\EndOfBibitem
\bibitem[Ji \emph{et~al.}(2018)Ji, Arbel, Zhang, Freudiger, Hou, Lin, Yang,
  Bacskai, and Xie]{intro20}
M.~Ji, M.~Arbel, L.~Zhang, C.~W. Freudiger, S.~S. Hou, D.~Lin, X.~Yang, B.~J.
  Bacskai and X.~S. Xie, \emph{Science Advances}, 2018, \textbf{4},
  eaat7715\relax
\mciteBstWouldAddEndPuncttrue
\mciteSetBstMidEndSepPunct{\mcitedefaultmidpunct}
{\mcitedefaultendpunct}{\mcitedefaultseppunct}\relax
\EndOfBibitem
\bibitem[Bachmann \emph{et~al.}(2006)Bachmann, Zezell, da~Costa~Ribeiro, Gomes,
  and Ito]{auto2006}
L.~Bachmann, D.~M. Zezell, A.~da~Costa~Ribeiro, L.~Gomes and A.~S. Ito,
  \emph{Applied Spectroscopy Reviews}, 2006, \textbf{41}, 575--590\relax
\mciteBstWouldAddEndPuncttrue
\mciteSetBstMidEndSepPunct{\mcitedefaultmidpunct}
{\mcitedefaultendpunct}{\mcitedefaultseppunct}\relax
\EndOfBibitem
\bibitem[Monici(2005)]{auto2005}
M.~Monici, \emph{Biotechnology Annual Review}, 2005, \textbf{11}, 227 --
  256\relax
\mciteBstWouldAddEndPuncttrue
\mciteSetBstMidEndSepPunct{\mcitedefaultmidpunct}
{\mcitedefaultendpunct}{\mcitedefaultseppunct}\relax
\EndOfBibitem
\bibitem[Richards-Kortum and Sevick-Muraca(1996)]{auto1996}
R.~Richards-Kortum and E.~Sevick-Muraca, \emph{Annual Review of Physical
  Chemistry}, 1996, \textbf{47}, 555--606\relax
\mciteBstWouldAddEndPuncttrue
\mciteSetBstMidEndSepPunct{\mcitedefaultmidpunct}
{\mcitedefaultendpunct}{\mcitedefaultseppunct}\relax
\EndOfBibitem
\bibitem[Moreno-García \emph{et~al.}(2018)Moreno-García, Kun, Calero, Medina,
  and Calero]{auto2018}
A.~Moreno-García, A.~Kun, O.~Calero, M.~Medina and M.~Calero, \emph{Frontiers
  in Neuroscience}, 2018, \textbf{12}, 464\relax
\mciteBstWouldAddEndPuncttrue
\mciteSetBstMidEndSepPunct{\mcitedefaultmidpunct}
{\mcitedefaultendpunct}{\mcitedefaultseppunct}\relax
\EndOfBibitem
\bibitem[Boyd(Academic Press, London, 2008)]{boyd}
R.~Boyd, \emph{\textit{Nonlinear Optics}}, Academic Press, London, 2008\relax
\mciteBstWouldAddEndPuncttrue
\mciteSetBstMidEndSepPunct{\mcitedefaultmidpunct}
{\mcitedefaultendpunct}{\mcitedefaultseppunct}\relax
\EndOfBibitem
\bibitem[Shen(Wiley-Interscience, New York, 2003)]{shen}
Y.-R. Shen, \emph{\textit{The Principles of Nonlinear Optics}},
  Wiley-Interscience, New York, 2003\relax
\mciteBstWouldAddEndPuncttrue
\mciteSetBstMidEndSepPunct{\mcitedefaultmidpunct}
{\mcitedefaultendpunct}{\mcitedefaultseppunct}\relax
\EndOfBibitem
\bibitem[Campagnola \emph{et~al.}(2002)Campagnola, Millard, Terasaki, Hoppe,
  Malone, and Mohler]{shg1}
P.~Campagnola, A.~Millard, M.~Terasaki, P.~Hoppe, C.~Malone and W.~Mohler,
  \emph{Biophysical Journal}, 2002, \textbf{82}, 493--508\relax
\mciteBstWouldAddEndPuncttrue
\mciteSetBstMidEndSepPunct{\mcitedefaultmidpunct}
{\mcitedefaultendpunct}{\mcitedefaultseppunct}\relax
\EndOfBibitem
\bibitem[Zipfel \emph{et~al.}(2003)Zipfel, Williams, Christie, Nikitin, Hyman,
  and Webb]{shg13}
W.~R. Zipfel, R.~M. Williams, R.~Christie, A.~Y. Nikitin, B.~T. Hyman and W.~W.
  Webb, \emph{Proceedings of the National Academy of Sciences}, 2003,
  \textbf{100}, 7075--7080\relax
\mciteBstWouldAddEndPuncttrue
\mciteSetBstMidEndSepPunct{\mcitedefaultmidpunct}
{\mcitedefaultendpunct}{\mcitedefaultseppunct}\relax
\EndOfBibitem
\bibitem[Green \emph{et~al.}(2017)Green, Delaine-Smith, Askew, Byers, Reilly,
  and Matcher]{shg14}
N.~H. Green, R.~M. Delaine-Smith, H.~J. Askew, R.~Byers, G.~C. Reilly and S.~J.
  Matcher, \emph{Scientific Reports}, 2017, \textbf{7}, 13331\relax
\mciteBstWouldAddEndPuncttrue
\mciteSetBstMidEndSepPunct{\mcitedefaultmidpunct}
{\mcitedefaultendpunct}{\mcitedefaultseppunct}\relax
\EndOfBibitem
\bibitem[Rivard \emph{et~al.}(2011)Rivard, Lalibert\'{e}, Bertrand-Grenier,
  Harnagea, Pfeffer, Valli\`{e}res, St-Pierre, Pignolet, Khakani, and
  L\'{e}gar\'{e}]{shg15}
M.~Rivard, M.~Lalibert\'{e}, A.~Bertrand-Grenier, C.~Harnagea, C.~P. Pfeffer,
  M.~Valli\`{e}res, Y.~St-Pierre, A.~Pignolet, M.~Khakani and
  F.~L\'{e}gar\'{e}, \emph{Biomedical Optics Express}, 2011, \textbf{2},
  26--36\relax
\mciteBstWouldAddEndPuncttrue
\mciteSetBstMidEndSepPunct{\mcitedefaultmidpunct}
{\mcitedefaultendpunct}{\mcitedefaultseppunct}\relax
\EndOfBibitem
\bibitem[Hanczyc \emph{et~al.}(2013)Hanczyc, Samoc, and Norden]{shg2}
P.~Hanczyc, M.~Samoc and B.~Norden, \emph{Nature Photonics}, 2013, \textbf{7},
  969–972\relax
\mciteBstWouldAddEndPuncttrue
\mciteSetBstMidEndSepPunct{\mcitedefaultmidpunct}
{\mcitedefaultendpunct}{\mcitedefaultseppunct}\relax
\EndOfBibitem
\bibitem[Tsai \emph{et~al.}(2004)Tsai, Grutzendler, Duff, and Gan]{shg7}
J.~Tsai, J.~Grutzendler, K.~J. Duff and W.-B. Gan, \emph{Nature Neuronscience},
  2004, \textbf{7}, 1181–1183\relax
\mciteBstWouldAddEndPuncttrue
\mciteSetBstMidEndSepPunct{\mcitedefaultmidpunct}
{\mcitedefaultendpunct}{\mcitedefaultseppunct}\relax
\EndOfBibitem
\bibitem[Rajamohamedsait and Sigurdsson(2012)]{shg4}
H.~B. Rajamohamedsait and E.~M. Sigurdsson, \emph{Histological Staining of
  Amyloid and Pre-amyloid Peptides and Proteins in Mouse Tissue. In: Sigurdsson
  E., Calero M., Gasset M. (eds) Amyloid Proteins}, Humana Press, 2012, vol.
  849, pp. 411--424\relax
\mciteBstWouldAddEndPuncttrue
\mciteSetBstMidEndSepPunct{\mcitedefaultmidpunct}
{\mcitedefaultendpunct}{\mcitedefaultseppunct}\relax
\EndOfBibitem
\bibitem[Fu \emph{et~al.}(2015)Fu, Wang, Psciuk, Xiao, Batista, and Yan]{shg8}
L.~Fu, Z.~Wang, B.~T. Psciuk, D.~Xiao, V.~S. Batista and E.~C.~Y. Yan,
  \emph{The Journal of Physical Chemistry Letters}, 2015, \textbf{6},
  1310--1315\relax
\mciteBstWouldAddEndPuncttrue
\mciteSetBstMidEndSepPunct{\mcitedefaultmidpunct}
{\mcitedefaultendpunct}{\mcitedefaultseppunct}\relax
\EndOfBibitem
\bibitem[Roeters \emph{et~al.}(2013)Roeters, van Dijk, Torres-Knoop, Backus,
  Campen, Bonn, and Woutersen]{shg9}
S.~J. Roeters, C.~N. van Dijk, A.~Torres-Knoop, E.~H.~G. Backus, R.~K. Campen,
  M.~Bonn and S.~Woutersen, \emph{The Journal of Physical Chemistry A}, 2013,
  \textbf{117}, 6311--6322\relax
\mciteBstWouldAddEndPuncttrue
\mciteSetBstMidEndSepPunct{\mcitedefaultmidpunct}
{\mcitedefaultendpunct}{\mcitedefaultseppunct}\relax
\EndOfBibitem
\bibitem[Tan \emph{et~al.}(2019)Tan, Zhang, Luo, and Ye]{shg12}
J.~Tan, J.~Zhang, Y.~Luo and S.~Ye, \emph{Journal of the American Chemical
  Society}, 2019, \textbf{141}, 1941--1948\relax
\mciteBstWouldAddEndPuncttrue
\mciteSetBstMidEndSepPunct{\mcitedefaultmidpunct}
{\mcitedefaultendpunct}{\mcitedefaultseppunct}\relax
\EndOfBibitem
\bibitem[Kurouski \emph{et~al.}(2015)Kurouski, Van~Duyne, and Lednev]{amideI1}
D.~Kurouski, R.~P. Van~Duyne and I.~K. Lednev, \emph{Analyst}, 2015,
  \textbf{140}, 4967--4980\relax
\mciteBstWouldAddEndPuncttrue
\mciteSetBstMidEndSepPunct{\mcitedefaultmidpunct}
{\mcitedefaultendpunct}{\mcitedefaultseppunct}\relax
\EndOfBibitem
\bibitem[Barth and Zscherp(2002)]{amideI2}
A.~Barth and C.~Zscherp, \emph{Quarterly Reviews of Biophysics}, 2002,
  \textbf{35}, 369–430\relax
\mciteBstWouldAddEndPuncttrue
\mciteSetBstMidEndSepPunct{\mcitedefaultmidpunct}
{\mcitedefaultendpunct}{\mcitedefaultseppunct}\relax
\EndOfBibitem
\bibitem[Kinney \emph{et~al.}(2018)Kinney, Bemiller, Murtishaw, Leisgang,
  Salazar, and Lamb]{alz1}
J.~W. Kinney, S.~M. Bemiller, A.~S. Murtishaw, A.~M. Leisgang, A.~M. Salazar
  and B.~T. Lamb, \emph{Alzheimer's \& Dementia: Translational Research \&
  Clinical Interventions}, 2018, \textbf{4}, 575 -- 590\relax
\mciteBstWouldAddEndPuncttrue
\mciteSetBstMidEndSepPunct{\mcitedefaultmidpunct}
{\mcitedefaultendpunct}{\mcitedefaultseppunct}\relax
\EndOfBibitem
\bibitem[Heneka \emph{et~al.}(2010)Heneka, O’Banion, Terwel, and
  Kummer]{alz2}
M.~Heneka, M.~O’Banion, D.~Terwel and M.~P. Kummer, \emph{Journal of Neural
  Transmission}, 2010, \textbf{177}, 919 – 947\relax
\mciteBstWouldAddEndPuncttrue
\mciteSetBstMidEndSepPunct{\mcitedefaultmidpunct}
{\mcitedefaultendpunct}{\mcitedefaultseppunct}\relax
\EndOfBibitem
\bibitem[Martin \emph{et~al.}(2017)Martin, Boucher, Fontaine, and
  Delarasse]{alz3}
E.~Martin, C.~Boucher, B.~Fontaine and C.~Delarasse, \emph{Aging Cell}, 2017,
  \textbf{16}, 27--38\relax
\mciteBstWouldAddEndPuncttrue
\mciteSetBstMidEndSepPunct{\mcitedefaultmidpunct}
{\mcitedefaultendpunct}{\mcitedefaultseppunct}\relax
\EndOfBibitem
\bibitem[Nagele \emph{et~al.}(2004)Nagele, Wegiel, Venkataraman, Imaki, Wang,
  and Wegiel]{alz5}
R.~G. Nagele, J.~Wegiel, V.~Venkataraman, H.~Imaki, K.-C. Wang and J.~Wegiel,
  \emph{Neurobiology of Aging}, 2004, \textbf{25}, 663 -- 674\relax
\mciteBstWouldAddEndPuncttrue
\mciteSetBstMidEndSepPunct{\mcitedefaultmidpunct}
{\mcitedefaultendpunct}{\mcitedefaultseppunct}\relax
\EndOfBibitem
\bibitem[Ahmed \emph{et~al.}(2010)Ahmed, Davis, Aucoin, Sato, Ahuja, Aimoto,
  Elliott, Van~Nostrand, and Smith]{phenyl1}
M.~Ahmed, J.~Davis, D.~Aucoin, T.~Sato, S.~Ahuja, S.~Aimoto, J.~I. Elliott,
  W.~E. Van~Nostrand and S.~O. Smith, \emph{Nature Structural \& Molecular
  Biology}, 2010, \textbf{17}, 561--567\relax
\mciteBstWouldAddEndPuncttrue
\mciteSetBstMidEndSepPunct{\mcitedefaultmidpunct}
{\mcitedefaultendpunct}{\mcitedefaultseppunct}\relax
\EndOfBibitem
\bibitem[Cukalevski \emph{et~al.}(2012)Cukalevski, Boland, Frohm, Thulin,
  Walsh, and Linse]{phenyl2}
R.~Cukalevski, B.~Boland, B.~Frohm, E.~Thulin, D.~Walsh and S.~Linse, \emph{ACS
  Chemical Neuroscience}, 2012, \textbf{3}, 1008--1016\relax
\mciteBstWouldAddEndPuncttrue
\mciteSetBstMidEndSepPunct{\mcitedefaultmidpunct}
{\mcitedefaultendpunct}{\mcitedefaultseppunct}\relax
\EndOfBibitem
\bibitem[Barth(2007)]{amideb1}
A.~Barth, \emph{Biochimica et Biophysica Acta (BBA) - Bioenergetics}, 2007,
  \textbf{1767}, 1073 -- 1101\relax
\mciteBstWouldAddEndPuncttrue
\mciteSetBstMidEndSepPunct{\mcitedefaultmidpunct}
{\mcitedefaultendpunct}{\mcitedefaultseppunct}\relax
\EndOfBibitem
\bibitem[Oberg \emph{et~al.}(2004)Oberg, Ruysschaert, and
  Goormaghtigh]{amideb2}
K.~A. Oberg, J.-M. Ruysschaert and E.~Goormaghtigh, \emph{European Journal of
  Biochemistry}, 2004, \textbf{271}, 2937--2948\relax
\mciteBstWouldAddEndPuncttrue
\mciteSetBstMidEndSepPunct{\mcitedefaultmidpunct}
{\mcitedefaultendpunct}{\mcitedefaultseppunct}\relax
\EndOfBibitem
\bibitem[Heinrich \emph{et~al.}(2008)Heinrich, Hofer, Ritsch, Ciardi, Bernet,
  and Ritsch-Marte]{unslip}
C.~Heinrich, A.~Hofer, A.~Ritsch, C.~Ciardi, S.~Bernet and M.~Ritsch-Marte,
  \emph{Optics Express}, 2008, \textbf{16}, 2699--2708\relax
\mciteBstWouldAddEndPuncttrue
\mciteSetBstMidEndSepPunct{\mcitedefaultmidpunct}
{\mcitedefaultendpunct}{\mcitedefaultseppunct}\relax
\EndOfBibitem
\bibitem[Borchelt \emph{et~al.}(1997)Borchelt, Ratovitski, van Lare, Lee,
  Gonzales, Jenkins, Copeland, Price, and Sisodia]{prep1997}
D.~R. Borchelt, T.~Ratovitski, J.~van Lare, M.~K. Lee, V.~Gonzales, N.~A.
  Jenkins, N.~G. Copeland, D.~L. Price and S.~S. Sisodia, \emph{Neuron}, 1997,
  \textbf{19}, 939 -- 945\relax
\mciteBstWouldAddEndPuncttrue
\mciteSetBstMidEndSepPunct{\mcitedefaultmidpunct}
{\mcitedefaultendpunct}{\mcitedefaultseppunct}\relax
\EndOfBibitem
\bibitem[Jankowsky \emph{et~al.}(2001)Jankowsky, Slunt, Ratovitski, Jenkins,
  Copeland, and Borchelt]{prep2001}
J.~L. Jankowsky, H.~H. Slunt, T.~Ratovitski, N.~A. Jenkins, N.~G. Copeland and
  D.~R. Borchelt, \emph{Biomolecular Engineering}, 2001, \textbf{17}, 157 --
  165\relax
\mciteBstWouldAddEndPuncttrue
\mciteSetBstMidEndSepPunct{\mcitedefaultmidpunct}
{\mcitedefaultendpunct}{\mcitedefaultseppunct}\relax
\EndOfBibitem
\bibitem[Garcia-Alloza \emph{et~al.}(2006)Garcia-Alloza, Robbins, Zhang-Nunes,
  Purcell, Betensky, Raju, Prada, Greenberg, Bacskai, and Frosch]{prep2006}
M.~Garcia-Alloza, E.~M. Robbins, S.~X. Zhang-Nunes, S.~M. Purcell, R.~A.
  Betensky, S.~Raju, C.~Prada, S.~M. Greenberg, B.~J. Bacskai and M.~P. Frosch,
  \emph{Neurobiology of Disease}, 2006, \textbf{24}, 516 -- 524\relax
\mciteBstWouldAddEndPuncttrue
\mciteSetBstMidEndSepPunct{\mcitedefaultmidpunct}
{\mcitedefaultendpunct}{\mcitedefaultseppunct}\relax
\EndOfBibitem
\end{mcitethebibliography}
\bibliographystyle{main}

\pagebreak
\widetext 
\noindent {\huge\textbf{Supplementary information for} \label{si}}
\\
\\
\\
\noindent\LARGE\textbf{Multimodal microscopy for characterization of amyloid-$\upbeta$ plaques biomarkers in animal model of Alzheimer's disease}
\\
\\

  \noindent\normalsize{Renan Cunha, Lucas Lafeta, Emerson A. Fonseca, Alexandre Barbosa, Marco A. Romano-Silva, Rafael Vieira, Ado Jorio, and Leandro M. Malard\textit{$^{\ast}$}} \\

\noindent$\ast$~E-mail: lmalard@fisica.ufmg.br.

\newpage

\begin{figure*}[tp!]
\setcounter{figure}{0}
\renewcommand\figurename{Fig. S}
\centering
\includegraphics[width=0.85\textwidth]{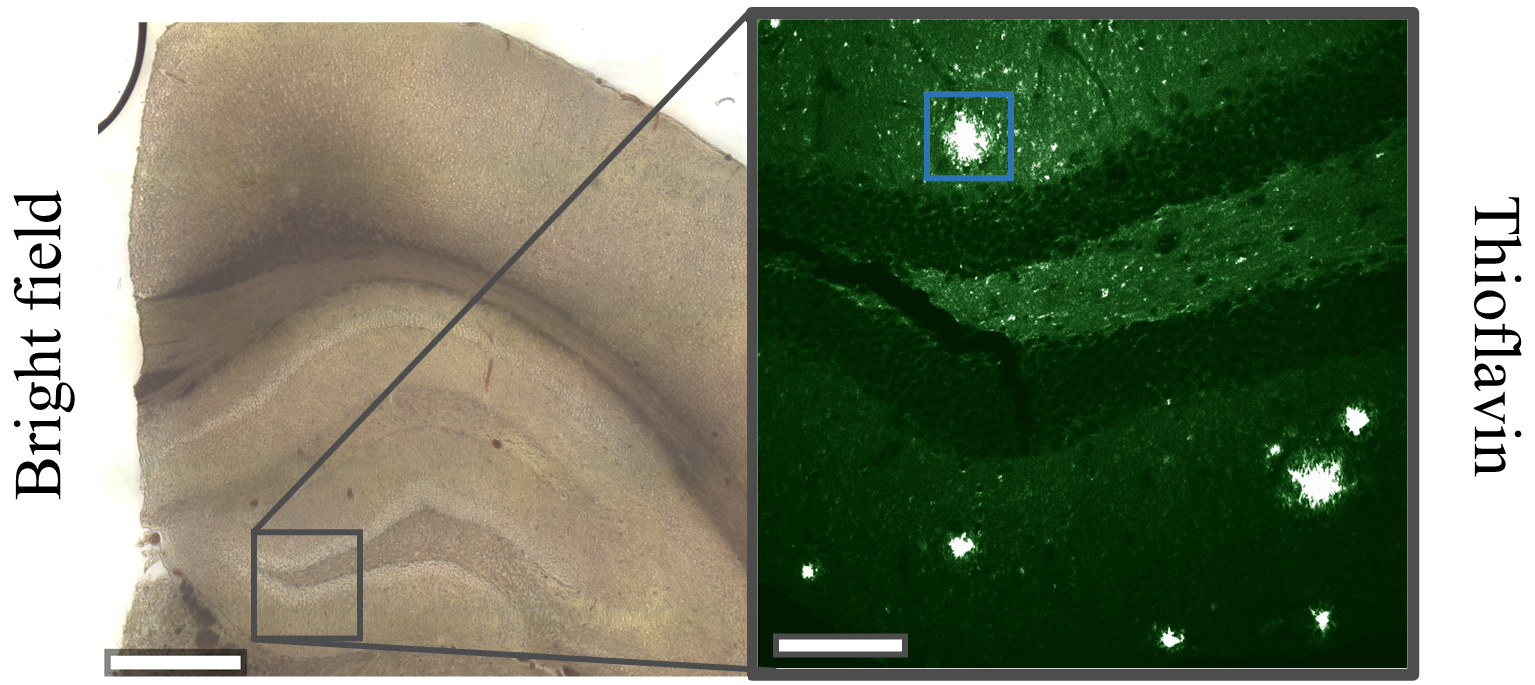}
\let\nobreakspace\relax
\caption{Tissue identification in the ThioS stained TPEF image. Same thioS stained TPEF image present in Fig. 1C of the main article but with saturated A$\upbeta$ plaques, favoring the identification of the hippocampus and showing other plaques in a 12-month-old animal. Left side: bright field image of the mouse brain slice (scale bar is 500 $\mu$m). Right side: ThioS stained TPEF image of the region shown by the grey square in the bright field image with several A$\upbeta$ plaques (scale bar is 100 $\mu$m). The blue square also identifies the plaque studied in Figure 1 of the manuscript.} \label{figs1}
\end{figure*}

\clearpage

\begin{figure*}[tp!]
\setcounter{figure}{1}
\renewcommand\figurename{Fig. S}
\centering
\includegraphics[width=0.75\textwidth]{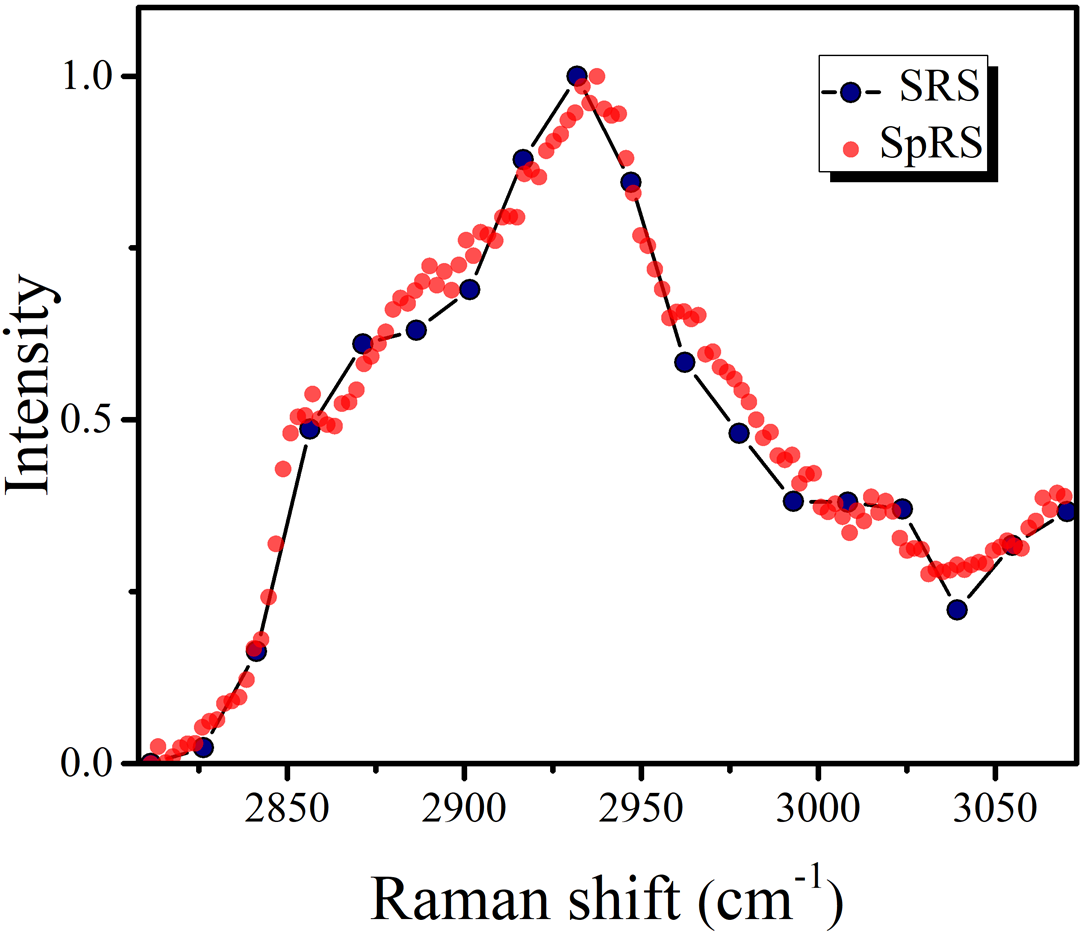}
\let\nobreakspace\relax
\caption{Stimulated Raman microscopy \textit{versus} spontaneous Raman spectra. Comparison between stimulated Raman scattering (SRS, blue) and spontaneous Raman scattering (SpRS, red) spectra obtained from mouse brain tissue images. The images were acquired from different scattering geometries (transmission for the SRS and reflection for the SpRS), but demonstrate a high spectral correspondence.}\label{figs2}
\end{figure*}

\clearpage

\begin{figure*}[tp!]
\setcounter{figure}{2}
\renewcommand\figurename{Fig. S}
\centering
\includegraphics[width=1\textwidth]{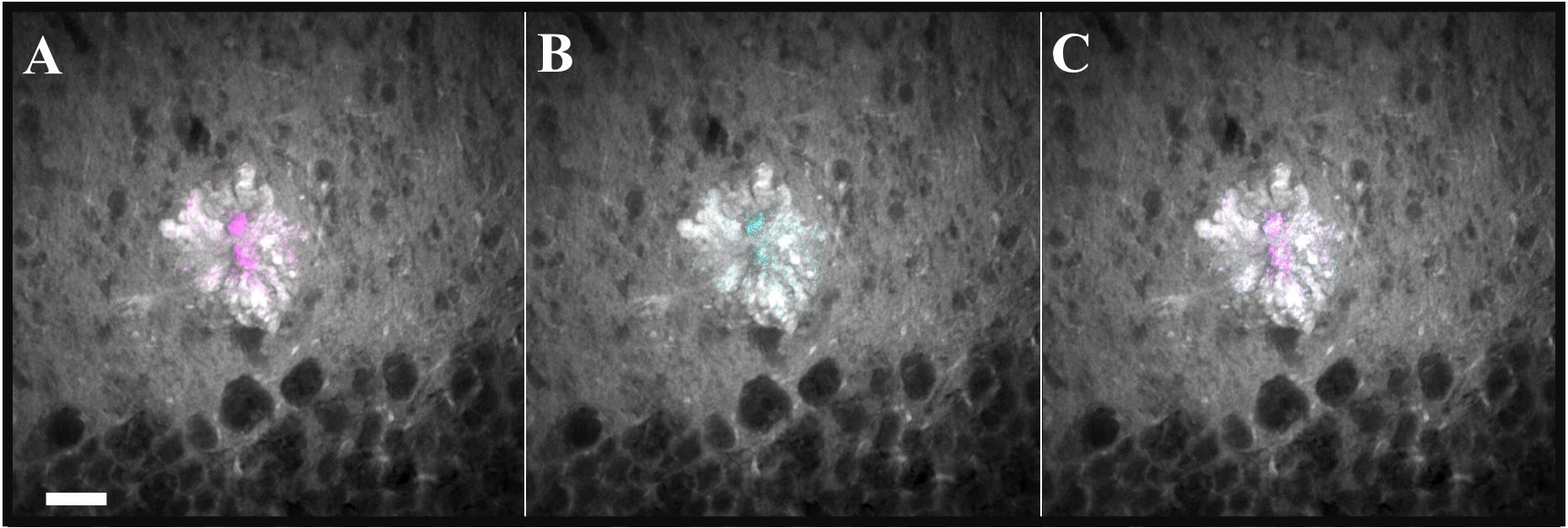}
\let\nobreakspace\relax
\caption{Comparison between high-resolution SRS images of amide I and phenylalanine and spatial distribution of both core biomarkers. (A,B) merged image of the halo and core, demonstrating the spatial distribution of the response based on the frequency of amide I (A) and phenylalanine (B). There is a spatial correlation in the distribution of both biomarkers in the core, although they present appreciable differences. (C) the merged image of amide I (A) and phenylalanine (B), as well as the halo, also demonstrate this distribution. In all figures, the halo image is based on the frequency attributed to lipids, 2850 cm$^{-1}$. Images without a scale bar share the scale bar of the leftmost image in the same line, which is 20 $\mu$m.}\label{figs3}
\end{figure*}

\clearpage

\begin{figure*}[tp!]
\setcounter{figure}{3}
\renewcommand\figurename{Fig. S}
\centering
\includegraphics[width=1\textwidth]{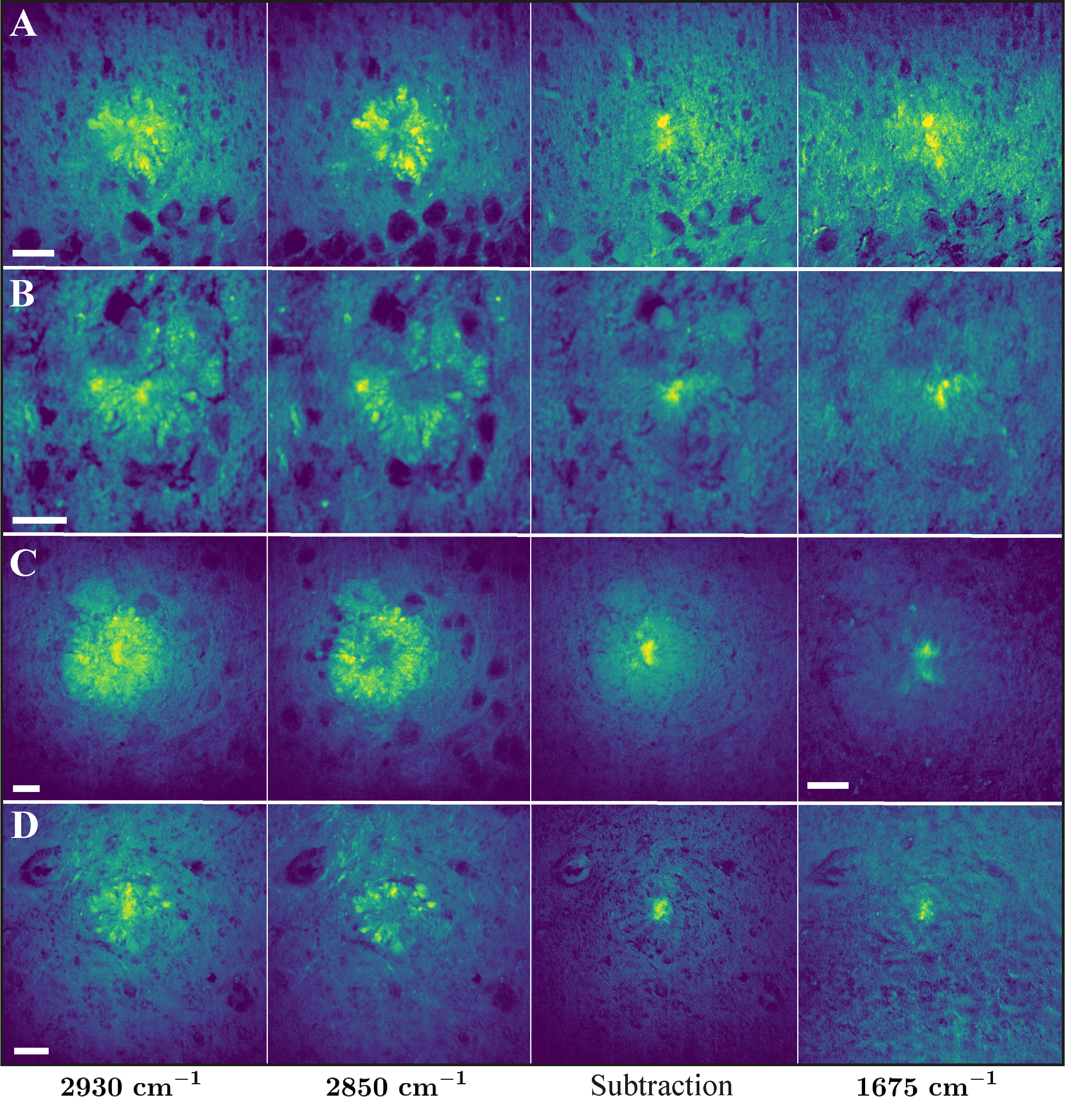}
\let\nobreakspace\relax
\caption{Subtraction of SRS images in the high-frequency region and comparison with SRS images based on the vibrational mode of amide I. (A-D) different A$\upbeta$ plaques distributed in three vibrations: 2930 (proteins/lipids), revealing both the core and the halo; 2850 (lipids), with only the halo in evidence; and 1675 (amide I), with only the core in evidence. This last column should be compared to the subtraction column obtained by subtracting images at frequencies 2930 and 2850 cm$^{-1}$. The images based on amide I and those obtained by subtraction show a significant spatial correlation.}\label{figs4}
\end{figure*}

\clearpage

\begin{figure*}[tp!]
\setcounter{figure}{4}
\renewcommand\figurename{Fig. S}
\centering
\includegraphics[width=1\textwidth]{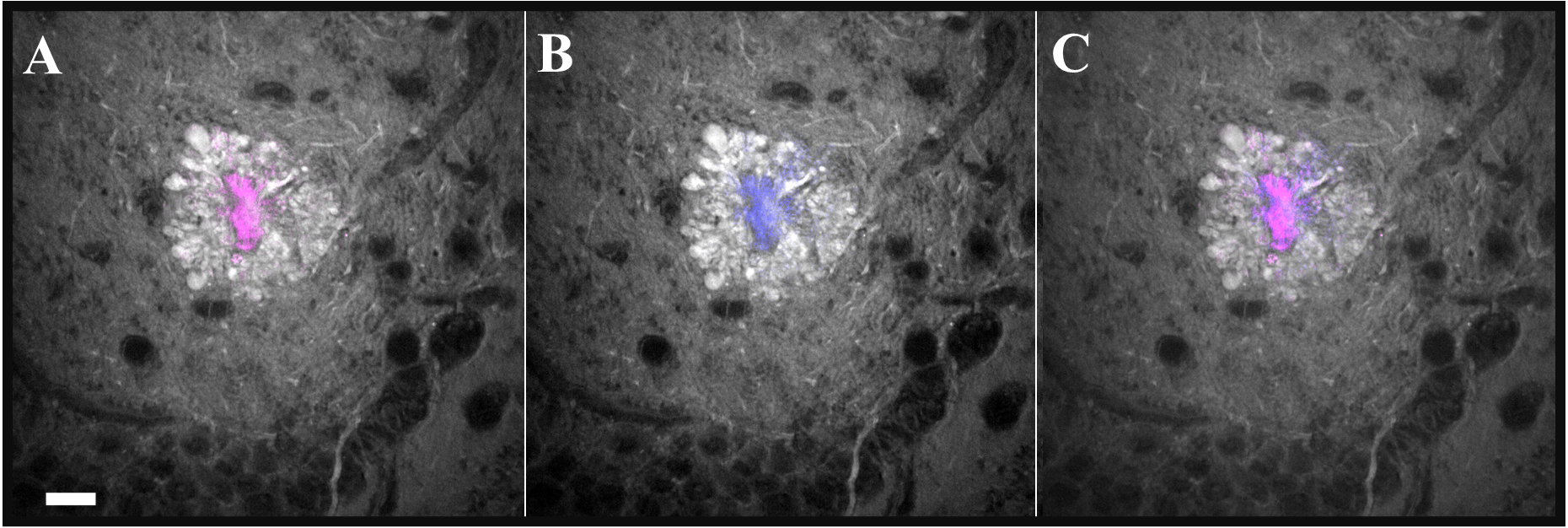}
\let\nobreakspace\relax
\caption{Comparison between high-resolution SRS images of amide I and amide B and spatial distribution of both core biomarkers. (A,B) merged image of the halo and core, demonstrating the spatial distribution of the response based on the frequency of amide I (A) and amide B (B). There is a spatial correlation in the distribution of both biomarkers in the core; however, no appreciable difference can be ensured. (C) the merged image of amide I (A) and amide B (B), as well as the halo, also demonstrate this distribution. In all figures, the halo image is based on the frequency attributed to lipids, 2850 cm$^{-1}$. Images without a scale bar share the scale bar of the leftmost image in the same line, which is 20 $\mu$m.}\label{figs5}
\end{figure*}

\clearpage

\begin{figure*}[tp!]
\setcounter{figure}{5}
\renewcommand\figurename{Fig. S}
\centering
\includegraphics[width=1\textwidth]{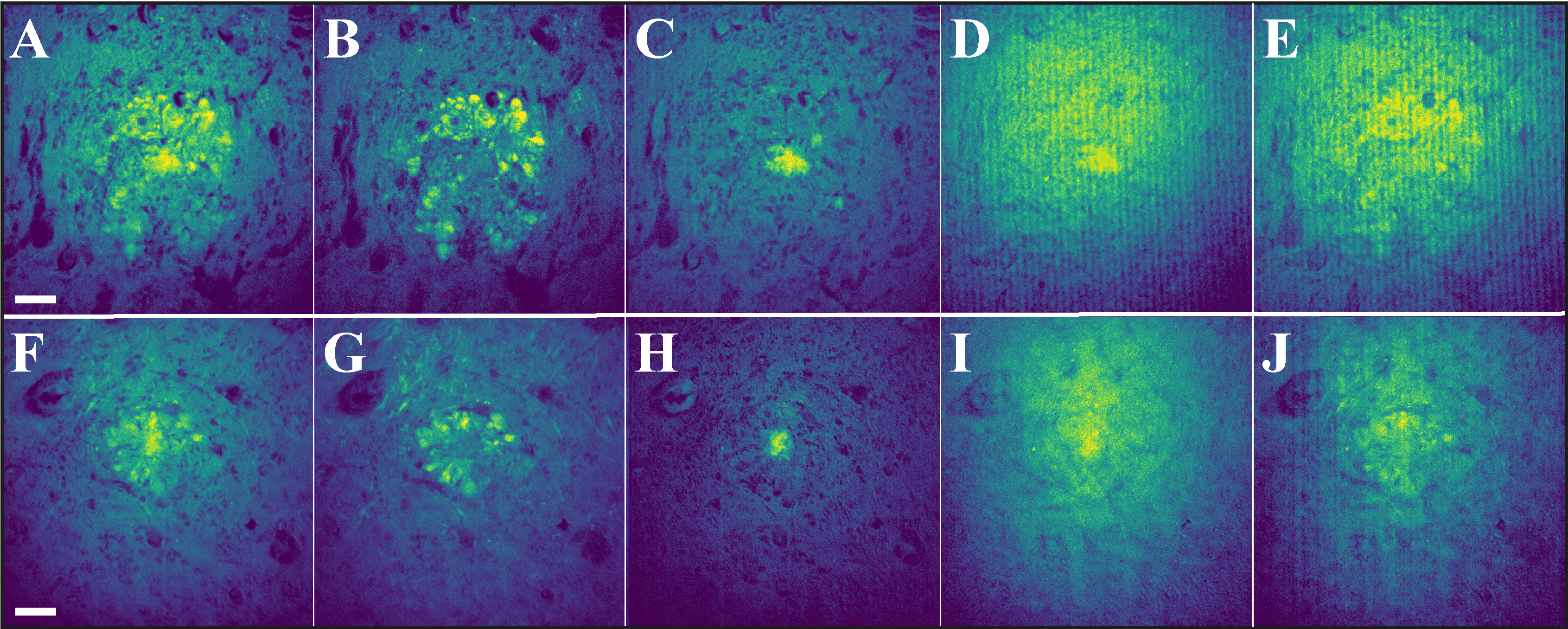}
\let\nobreakspace\relax
\caption{High-resolution SRS image with core and halo biomarkers in the high-frequency spectral region for two different A$\upbeta$ plaques.(A,F) and (B,G) are images based on the known  2930 cm$^{-1}$ (proteins/lipids) and 2850 cm$^{-1}$ (lipids) frequencies, respectively. (C,H) is the subtraction of these images, showing the core in the high-frequency region. (D,I) is the image based on the 3070 cm$^{-1}$ (amide B) frequency, which shows the core in the high-frequency region, correlating with the image obtained by subtraction (C,H). (E,J) is the halo image based on the 3019 cm$^{-1}$ frequency attributed to unsaturated lipids. Images without a scale bar share the scale bar of the leftmost image in the same line, which is 20 $\mu$m.
}\label{figs6}
\end{figure*}

\end{document}